\documentclass[twocolumn,epjc3]{svjour3}
\smartqed  
\RequirePackage{graphicx}

\usepackage{color}
\usepackage{cite}

\begin{document}
\title{Optical properties of Kerr-Newman spacetime in the presence of plasma}

\author{
Gulmina Zaman Babar\thanksref{1,a} \and
Abdullah Zaman Babar\thanksref{2,b} \and
Farruh Atamurotov\thanksref{3,c,d} \and}
\institute{School of Natural Sciences, National University of Sciences and Technology, Sector H-12, Islamabad, Pakistan\label{a}\and Department of Electrical Engineering, Air University, Islamabad, Pakistan\label{b}\and
Inha University in Tashkent, Ziyolilar 9, Tashkent 100170, Uzbekistan\label{c}\and
 Ulugh Beg Astronomical Institute, Astronomy St. 33, Tashkent 100052, Uzbekistan\label{d}}
\thankstext{1}{\emph{e-mail:} gulminazamanbabar@yahoo.com}
\thankstext{2}{\emph{e-mail:} abdullahzamanbabar@yahoo.com}
\thankstext{3}{\emph{e-mail:} atamurotov@yahoo.com, fatamurotov@gmail.com}


\maketitle
\begin{abstract}
We have studied the null geodesics in the background of the Kerr-Newman black hole veiled by a plasma medium
using the Hamilton-Jacobi method. The influence of black hole's charge and plasma parameters on the
effective potential and the generic photon orbits has been investigated. Furthermore, our discussion embodies
the effects of black hole's charge, plasma and the inclination angle on the shadow cast by the gravity with
and without the spin parameter. We examined the energy released from the
black hole as a result of the thermal radiations, which exclusively depends on
the size of the shadow. The angle of deflection of the massless particles is also explored considering a weak-field approximation.
We present our results in juxtaposition to the analogous black holes in General Relativity, particularly the Schwarzschild and Kerr black hole.
\end{abstract}

\maketitle
\section{Introduction} \label{intro}
The existence of super massive black holes has been investigated extensively for nearly two decades, through various esoteric astrophysical phenomena. Recently, The Event Horizon Telescope(EHT) project has been observed first direct image of M87* black hole\cite{aki:2019a,aki:2019b} using very long baseline interferometer(VLBI).
The physical structure of black holes is well apprehended by the shadow imaged by it, which is created when the black hole confronts a luminous source.
 Synge \cite{Synge:1966a} studied the shadow of the Schwarzschild black hole, which was then termed as the “escape cones” of light.
The radius of the shadow was calculated in terms of mass of the black hole and the radial coordinate where the observer is located.
Unlike a static black hole, the shadow of a rotating black hole is not a circular disk.
The first, foremost accurate calculations of the shadow were done by Bardeen considering
the Kerr space time \cite{Bar:1973a}. So far, the latter feature of the black hole has been widely investigated for various
gravities adopting a similar approach using classical method \cite{Hio:2009a,amr:2010a,Far:2013a,Wei:2013a,Fara:2013a,Fara:2013b,Fara:2014a,Fara:2014b,
Gre:2014a,Li:2014a,Gre:2015b,Li:2015a,Abu:2016b,Fara:2016a,Fara:2016b,Bis:2018a,stuch:2018a,Wei:2019a,Kum:2019a,Kum:2020a}.

 The influence of plasma medium on the events taking place in the black hole
 vicinity contributes an additional insight into its physical properties. The relativistic effects of plasma tracing
 light rays in the surroundings of compact objects are thoroughly studied in \cite{Ave:2003a}. A detailed discussion
 about gravitational lensing in the presence of a non-uniform plasma is carried out by Bisnovatyi-Kogan and Tsupko in
 \cite{Bin:2010a}. Later on, they extended their research for the Schwarzschild spacetime \cite{Tsp:2011a,Tsp:2014a,Tsp:2015a}.
 One may get specific details from \cite{Abu:2013a,Rog:2015a,Far:2016a,Bis:2017a,Abu:2017aa,Abu:2017a,Car:2018a} in
 reference to the above mentioned analysis.

Nowadays, shadow of the black hole in the presence of plasma has become the field of interest for researchers. Recently,
 a profound examination has been established to study the shadow of the Schwarzschild and Kerr space-time coupled with a plasma medium
 in the following papers \cite{Per:2015a,Abu:2015a} using the Synge formulism  \cite{Synge:1960b} and the performance of the plasma medium work was studied using a different approach in \cite{Tsp:2017a}. We shall put forth the Synge formulism analysis
 in analogy to the aforementioned papers to retrace the influence of plasma on the Kerr-Newman space-time. It is a stationary and an
axisymmetric solution to the Einstein-Maxwell equations depending on the mass, angular
momentum and electrical charge of the black hole.
The surface geometry of the Kerr-Newman metric and its
 physical properties are well described in \cite{Neo:1981a}. After this work was published, several works were performed in a charged black hole
  \cite{Mad:1985aa,Mad:1987ab,Hod:1987ac}.

The rest of our paper is organized as follows. In Sec.~(\ref{metric}), we consider
the equations of motion of photons around an axially
symmetric black hole in the presence of a plasma. In Sec.~(\ref{effphoton}) the
effective potential along with the generic photon orbits are studied. Formalism for
the shadow cast by the space-time under consideration is
set-up in Sec.~(\ref{Sh1}). The subsections of  Sec.~(\ref{Sh1}) incorporate the analysis of the shadow and energy
emission by taking into account a non rotating charged black hole.
Sec.~(\ref{lensing}) includes an elaborate analysis of
the deflection angle caused by the deviation of photons in a weak-field approximation.
Finally, in Sec.~(\ref{con}) we summarize our main results.

\section{Photon Motion Around the Charged Black hole in the Presence of a Plasma}\label{metric}
\begin{figure}
\centering
\includegraphics[scale=0.42]{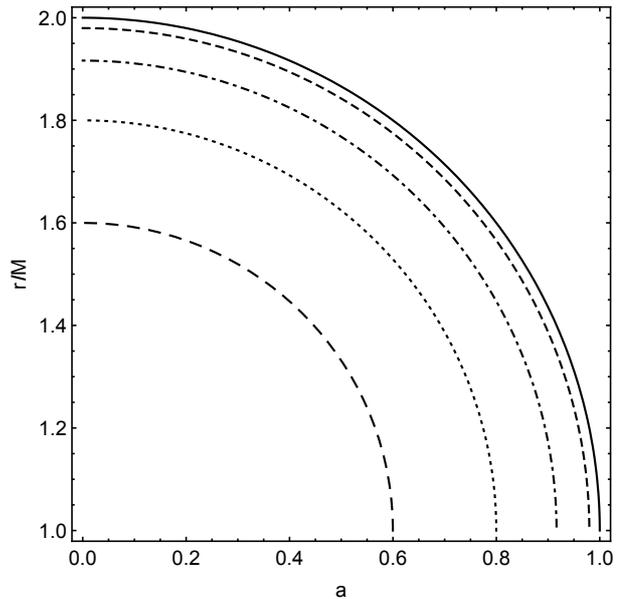}
\caption{The spin parameter $a$ dependence of the radial coordinate $r$ for
the different values of electric charge $Q$. The values assigned to $Q$ (top to bottom) are 0,0.2,0.4,0.6 and 0.8.}\label{event}
\end{figure}

In Boyer-Lindquist coordinates the charged rotating Kerr-Newman spacetime, an exact solution of the
Einstein-Maxwell field equations, is characterized by the line element \cite{Mad:1985aa,Mad:1987ab,Hod:1987ac},
\begin{eqnarray}
ds^2&=&-\frac{\Delta}{\rho^2}(dt-a \sin^2\theta d\phi)^2+\frac{\rho^2}{\Delta} dr^2+\rho^2 d\theta^2
\nonumber \\ &&+\frac{\sin^2\theta}{\rho^2}[a dt-(r^2+a^2) d\phi]^2,\label{Knmetric}
\end{eqnarray}
\par

where $\Delta=r^2-2Mr+a^2+Q^2$ and $\rho^2=r^2+a^2\cos^2\theta$. The parameters $M$, $Q$ and $a$ corresponds to the total mass, electric charge and
 specific angular momentum of the black hole, respectively. The Kerr metric is recovered when the charge $Q$ is annulled. Roots of the function $\Delta$,
\begin{equation}
r\pm= M\pm \sqrt{M-a^2-Q^2},
\end{equation}

determine the radii of the inner and outer horizons of the black hole. The inner horizon $r_{-}$ and the
 outer horizon $r_{+}$ are generally termed as the \textit{Cauchy horizon} and the \textit{event horizon}, respectively.
 The event horizon acts as a threshold from where no turning back is possible. The charge $Q$ has an evident
 influence on the horizon of the black hole which moves to a farther position at $Q=0$, shown in Fig. (\ref{event}).

We consider a static inhomogeneous plasma in the gravitational field with a refractive index $n$,
the expression of which was formulated in general terms by Synge \cite{Synge:1960b},

\begin{eqnarray}
n^2&=&1+\frac{p_{\mu} p^{\mu}}{(p_{\nu}u^{\nu})^2},
\end{eqnarray}

$p_{\mu}$ and $u^{\nu}$ refers to the four-momentum and four-velocity of the massless particle. One may obtain
 the vacuum case when $n=1$. The Hamilton-Jacobi equation for a black hole surrounded by a plasma is

\begin{eqnarray}
H(x^{\mu},p_{\mu})&=&\frac{1}{2}\bigg[g^{\mu\nu}p_{\mu}p_{\nu}+(n^2-1)(p_{\nu}u^{\nu})^2\bigg].
\end{eqnarray}
Now we use the Hamilton-Jacobi equation
which defines the equation of motion of the photons for a given space-time geometry

\begin{eqnarray}
H(x^{\mu},p_{\mu})&=&\frac{1}{2}\bigg[g^{\mu\nu}p_{\mu}p_{\nu}-(n^2-1)\bigg(p_{0}\sqrt{-g^{00}}\bigg)^2\bigg].\label{Hamilton}
\end{eqnarray}

The equations $\dot{x^{\mu}}=\partial H/\partial p_{\mu}$ and $\dot{p_{\mu}}=\partial H/\partial x^{\mu}$
define the trajectories of the photon in the plasma medium, given as below

\begin{eqnarray}
\rho^2 \dot{t}&=&\frac{(a^2+r^2)}{\Delta}\big(n^2E(a^2+r^2)-aL\big)
\nonumber \\ &&+a\sin^2\theta\bigg(\frac{L}{\sin^2\theta}-an^2E\bigg), \label{motiona}\\
 \rho^2\dot{\phi}&=&\frac{a}{\Delta}(E(a^2+r^2)-aL)+\bigg(\frac{L}{\sin^2\theta}-aE\bigg),\label{motionb}\\
 \rho^2\dot{r}&=&\sqrt{R},\label{motionc}\\
 \rho^2\dot{\theta}&=&\sqrt{\Theta}.\label{motiond}
\end{eqnarray}

The overdot denotes differentiation with respect to the particle’s proper time $\tau$. The functions $R(r)$ and $\Theta(\theta)$
 admit the following expressions,
\begin{eqnarray}
R&=&-\Delta(\mathcal{K}-2 aLE)+\big(nE(a^2+r^2)-aL\big)^2
\nonumber \\ &&+2aLE(n-1)(a^2+r^2),\label{Rorbit} \\
\Theta&=&\mathcal{K}-\frac{1}{\sin^2\theta}\big(L^2+a^2n^2E\sin^4\theta\big).
\end{eqnarray}

Here, $\mathcal{K}$ is the constant of separation. $E$ and $L$ are the conserved quantities acting along the
axis of symmetry termed, respectively, as the energy and angular momentum of the photon.

\section{Effective Potential and Photon Sphere}\label{effphoton}

For a systematic rational reasoning, it is necessary to introduce a specific form of the refractive index $n$ \cite{Bin:2010a,Tsp:2011a}, defined as follows
\begin{eqnarray}
n^2&=&1-\frac{{\omega_{e}}^2}{\omega^2}.\label{refractiveindex}
\end{eqnarray}
Here, $\omega_e$ is the plasma electron frequency and $\omega$ is the photon frequency perceived by a distant observer. The photon frequency depends on
the spatial coordinates $x^\mu$ due to the gravitational field. The light propagation through
the plasma medium is possible, provided that $\omega^2>{\omega_{e}}^2$. The plasma frequency has the following analytic expression

 \begin{eqnarray}
 \omega_{e}&=&\frac{4\pi e^2 N(r)}{m},
 \end{eqnarray}

where $e$, $m$ and $N(r)$ are the charge, mass and number density of the electron, respectively. By the implication of radial law density \cite{Rog:2015a}

\begin{eqnarray}
N(r)&=&\frac{N_{0}}{r^{h}},
\end{eqnarray}

where $h\geq0$, the plasma frequency becomes

\begin{eqnarray}
{\omega_{e}}^2&=&\frac{k}{r^h}.
\end{eqnarray}

For simplicity, we shall take $h=1$\cite{Rog:2015a,Abu:2015a}.
The radial potential can be directly evaluated from (\ref{motionc}) to study the generic photon behaviour in the presence of plasma,

\begin{eqnarray}
V_{\mathrm{eff}}&=&-\frac{a^2\big( L^2 -aE (2L -an^2E )\big)+\Delta(2aLE-\mathcal{K})}{r^4}
\nonumber \\ &&+\frac{2 a E (L -a n^2 E)-n^2E^2r^2}{r^2}.
\end{eqnarray}

\begin{figure*}
\includegraphics[width=.45\textwidth]{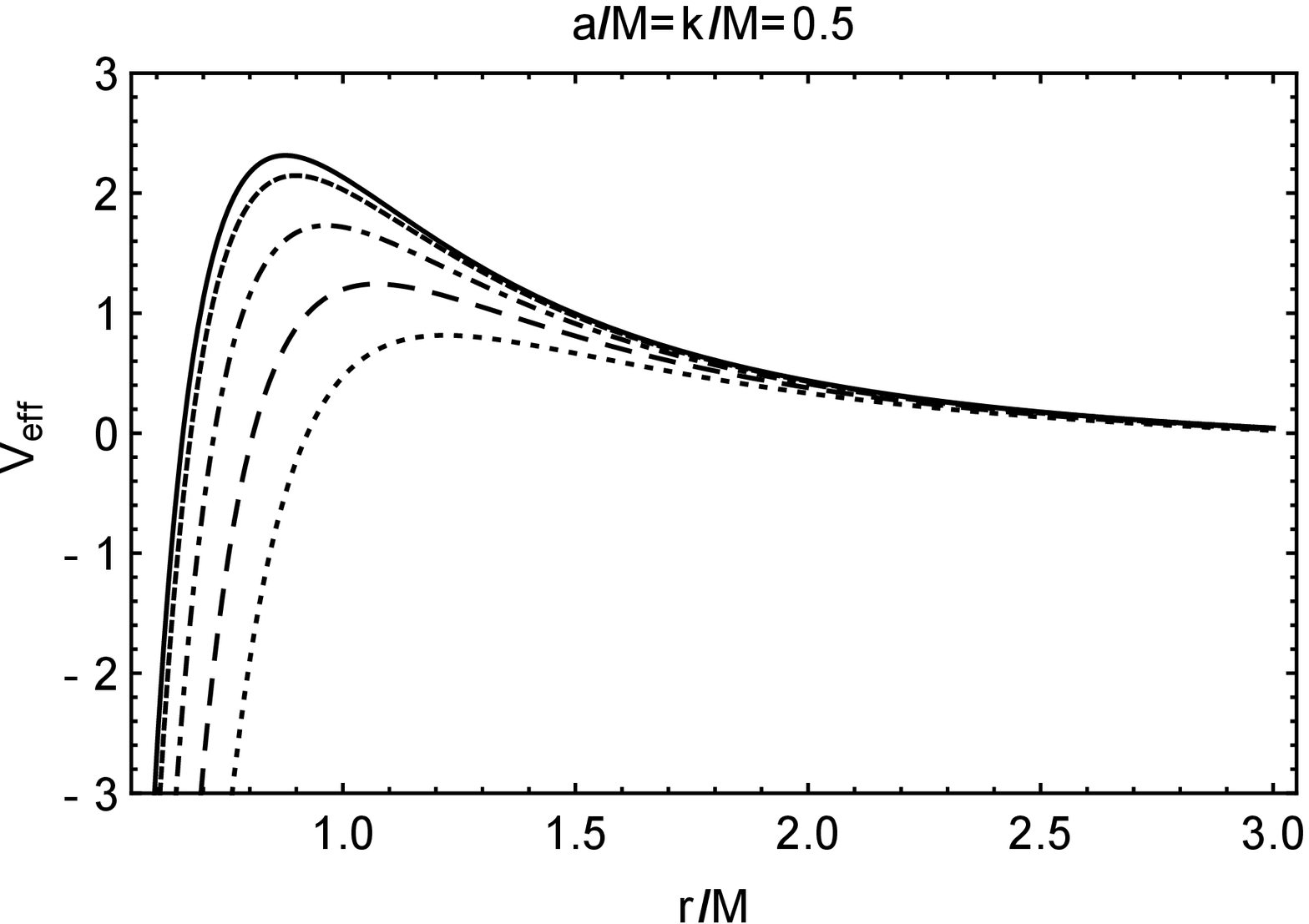}
\includegraphics[width=.45\textwidth]{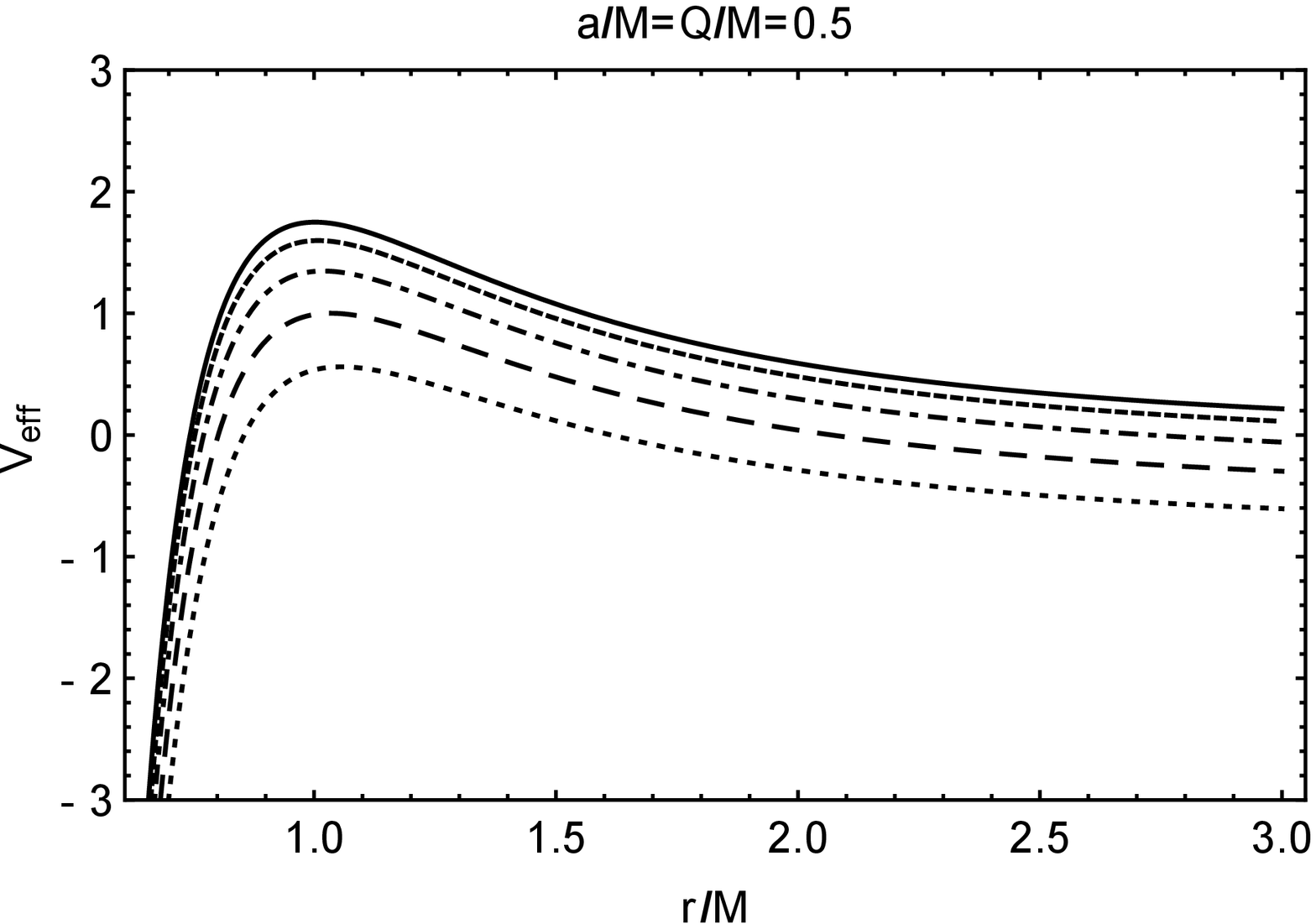}
\caption{Radial potential for photons with fixed parameters $E=0.9$ and $L=4$. In the left panel
the values ascribed to $Q$ (top to bottom) are 0,0.2.0.4,0.6 and 0.8 and in the right panel $n$ (top to bottom)
 has the values 0.2,0.4,0.6,0.8 and 1.}\label{Veff1}
\end{figure*}

 The circular orbits exist at $\dot{r}=0$ and the massless particle attains its stability at a
 fixed stationary position $r$ under the constraint $\partial_{r}V_{\mathrm{eff}}$. The particle may follow
 a marginally stable circular motion between the relative extrema satisfying the condition $\partial_{r}^2V_{\mathrm{eff}}$ \cite{Babar:2017a}.
The charge $Q$ of the black hole gives an additional strength to it, as illustrated in  Fig.(\ref{Veff1}).
For the Kerr Black hole at $Q$=0, the effective potential attains its maximum value with the highest potential barrier.
It is observed that the minimum values of $Q$ yields an extra shield to the photons against the black hole gravity,
hence, enhancing their stability. The plasma density increases with an increase in the refractive index,
consequently, the photon's strength depletes to carry on its motion in the black hole vicinity.
 It is also incurred that the barrier is sufficiently reduced when $n=1$ (vacuum case) \cite{Abu:2015a}.

We follow the formulism of \cite{Per:2015a} to find the photon orbits,
by considering a pressureless inhomogeneous plasma in the equatorial plane, $\theta=\pi/2$. Using (\ref{Hamilton})
we obtain the flight path of the photons in a unique way,

\begin{eqnarray}
\bigg(\frac{dr}{d\phi}\bigg)^2&=&\frac{-g^{rr}(g^{tt}E^2+g^{\phi\phi}L^2-2g^{t\phi}LE+{w_{e}}^2)}{(g^{\phi\phi}L-g^{t\phi}E)^2}.
\end{eqnarray}

\begin{figure*}
 \begin{center}
  \includegraphics[width=.45\textwidth]{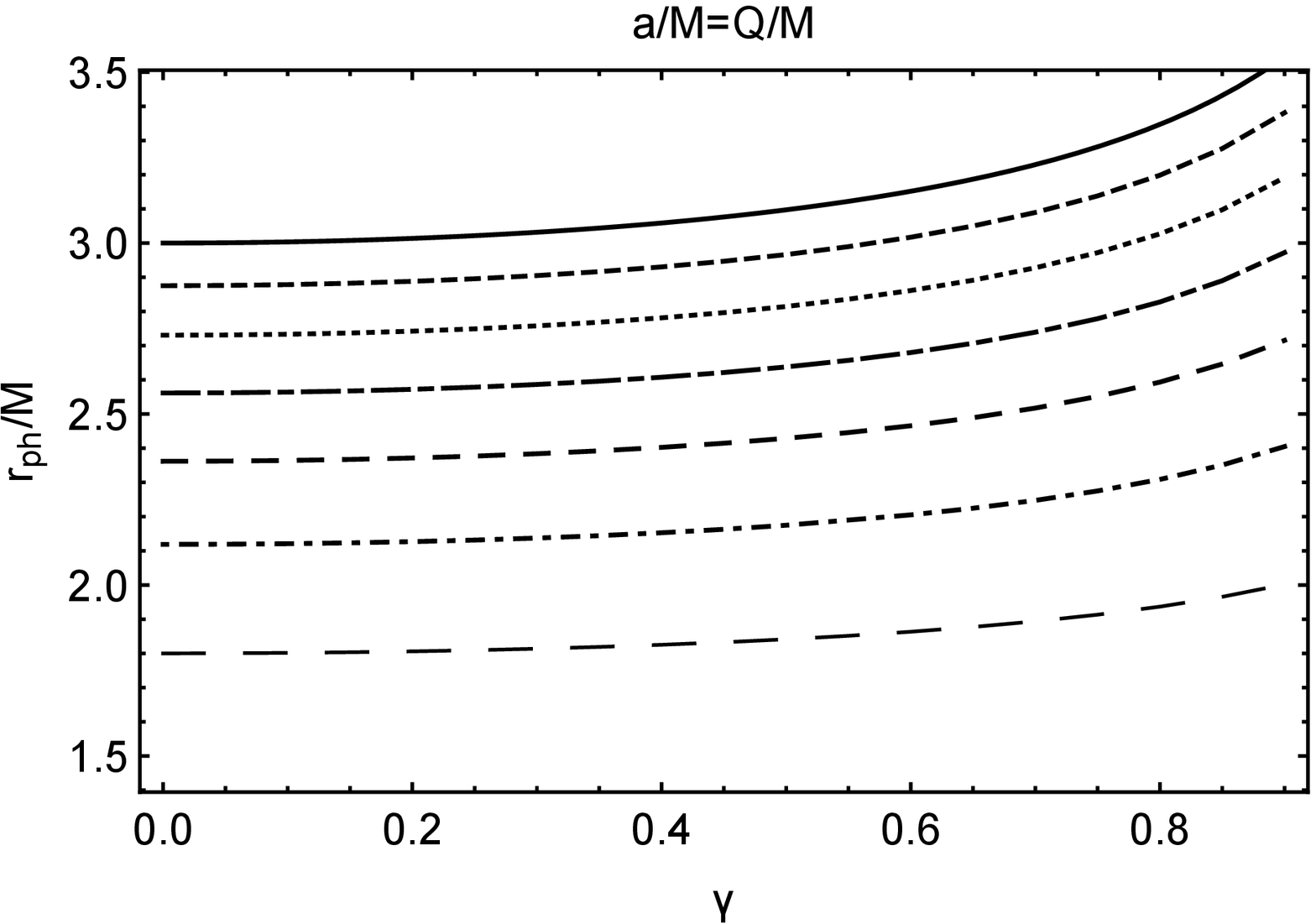}
   \includegraphics[width=.45\textwidth]{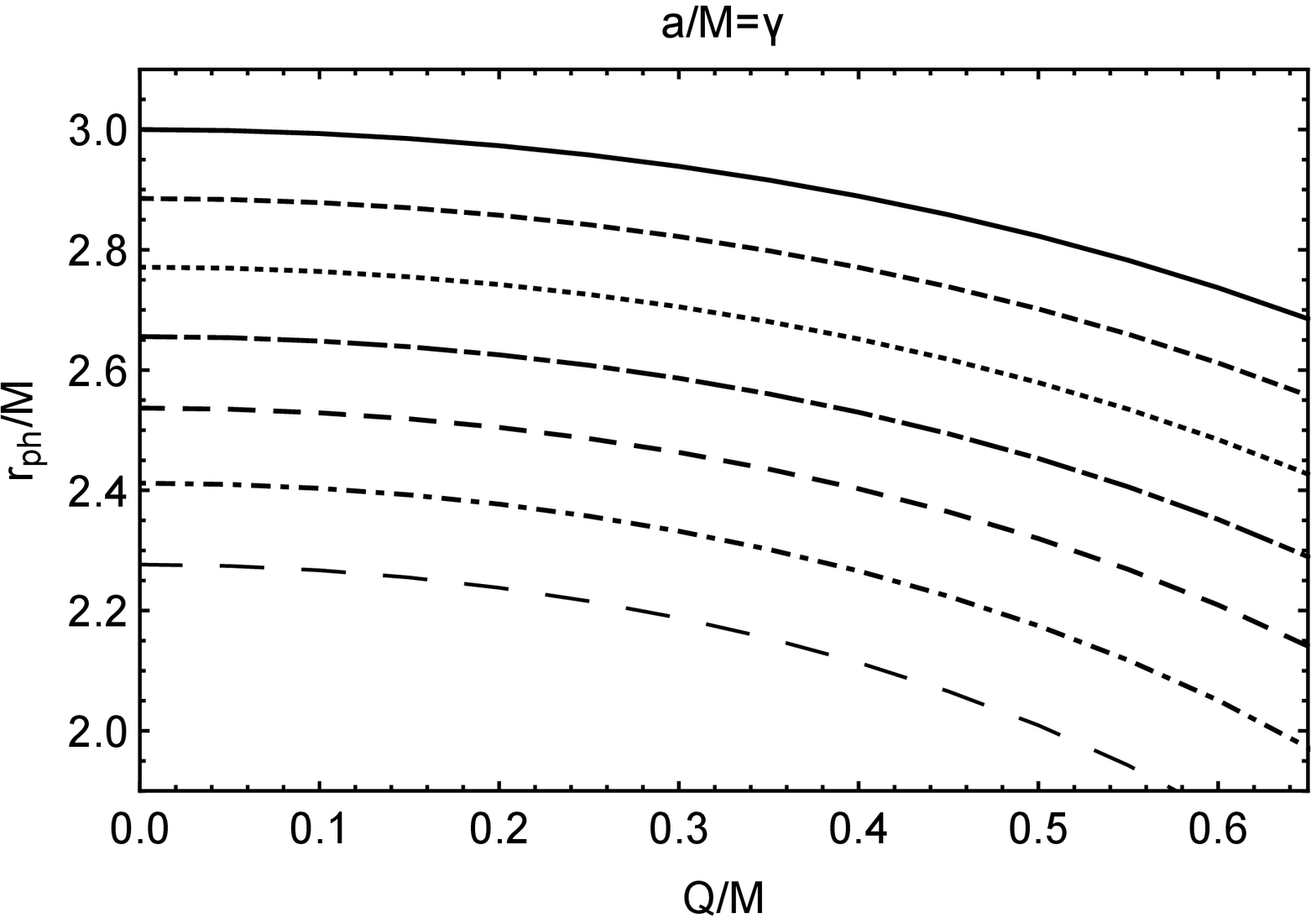}
  \end{center}
\caption{The photon orbits by varying $\gamma$ and $Q$. In the left and right panel from top to bottom the values of $Q$ and $\gamma$  are 0,0.1,0.2,0.3,0.4,0.5 and 0.6.}\label{Rph}
\end{figure*}

For a circular photon orbit $r_{\mathrm{ph}}$ we have $dr/d\phi$=0, which further yields a distinguished special parameter $L/E$ in the form of a function

\begin{eqnarray}
\xi(r)=\frac{L}{E}&=&\frac{g^{t\phi}+\sqrt{{(g^{t\phi})^2-g^{\phi\phi}}(g^{tt}+\frac{{w_e}^2}{E^2})}}{g^{\phi\phi}},
\end{eqnarray}

here, $\gamma=\frac{w_e}{E}$ is the dimensionless plasma constant. Note that $\gamma$=0 corresponds to the vacuum case.
 The working out of $d \xi(r)/d r=0$ gives a general expression, the roots of which provide the radius of the photon orbits,
\begin{eqnarray}
0&=&(\Delta -a^2)\big[r^2 \big(2 Q^2+r(-3M+r)\big)-\gamma^2\big(\Delta-a^2\big)^2\big]-
\nonumber \\ &&a(Q^2-M r\big)\big[2r\sqrt{\Delta(r^2-\gamma^2(\Delta-a^2))}+
 \nonumber \\ &&a((\Delta-a^2)\gamma^2-2r^2)\big].\label{photono}
\end{eqnarray}
We have solved numerically (\ref{photono}) to examine the influence of plasma and charge parameter
on the spherical photon orbits. The left panel of Fig.(\ref{Rph}) exhibits the dependence of the radius
on the plasma parameter for different values of $Q$. The $r_{\mathrm{ph}}$ increases in the presence of $\gamma$, hence,
the orbit shifts at a far distance due to rise in the plasma factor. The right panel of Fig.(\ref{Rph}) shows the effect of charge parameters
on the radius for different values of $\gamma$. Thus, it is inferred that the orbits come closer to the black hole due to the electric field intensity.
It is noticed that the radius of the orbits is  generally larger at $Q$=0. It is worth mentioning that the silhouette of the black hole,
elaborated in the subsequent section is observed only when $r_{\mathrm{ph}}>r_{+}$ \cite{Hio:2009a}, i.e, the radius of the spherical photon
orbit must be greater than that of the event horizon. An infinitesimal gravitational perturbation would drive the massless particles into
the black hole or toward spatial infinity.

\section{Shadow of Kerr-Newman Black hole in Presence of Plasma}\label{Sh1}

We first present an ansatz for the analysis of the shadow cast by the Kerr-Newman gravity in the presence of a plasma.
Let us consider the black hole between a bright source of light and an observer situated at fixed a Boyer-Lindquist coordinate
 $(r_{0},\theta_{0})$, where $\theta_{0}$ is the inclination angle between the rotation axis of the black hole and the line of
  sight of the observer and withal $r_{0}\rightarrow\infty$. The light waves emitted by the source reach the observer after gravitational
 deflection whereas, the photons with comparably less impact factor gets absorbed by the black hole. As a result, a dark patch in
 the space is created which is called the \textit{shadow}. The boundary of this shadow provides us details regarding the intrinsic
  configuration of the black hole. Utilizing the conserved quantities along with the constant $\mathcal{K}$, one can conveniently
  introduce the impact parameters $\xi=\frac{L}{E}$ and $\eta=\frac{\mathcal{K}}{E^2}$. With reference to (\ref{Rorbit})
  the orbits must satisfy the conditions, $R(r)=\partial_{r}R(r)=0$ which are fulfilled by the impact parameters and are further
  used to explore the contour of the black hole shadow.

\begin{figure*}
 \begin{center}
   \includegraphics[width=.23\textwidth]{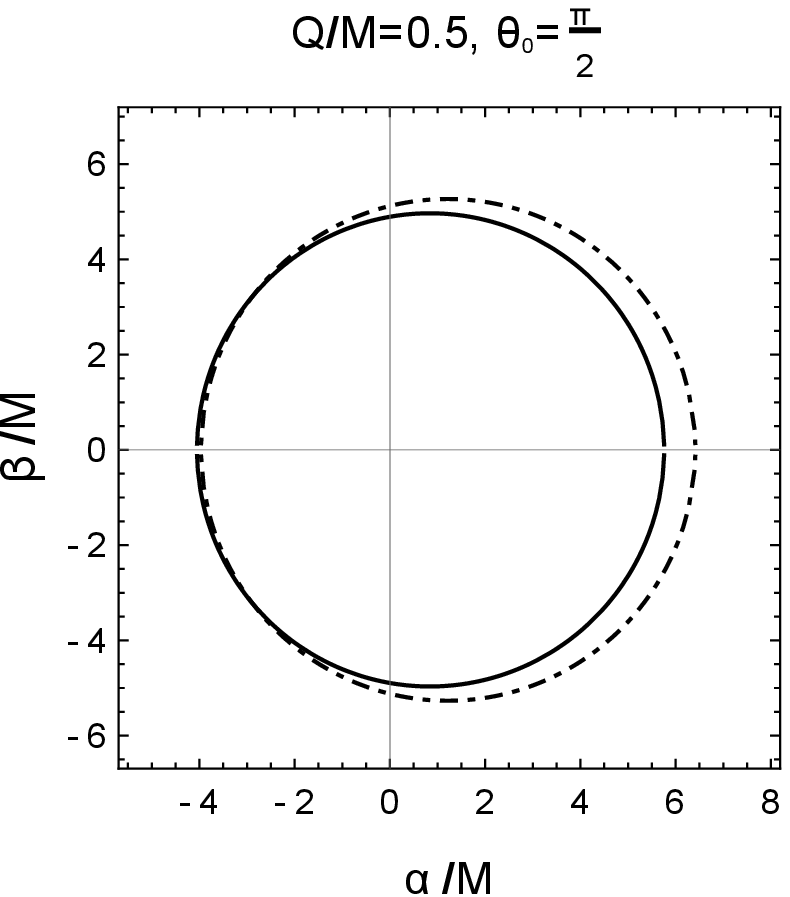}
   \includegraphics[width=.23\textwidth]{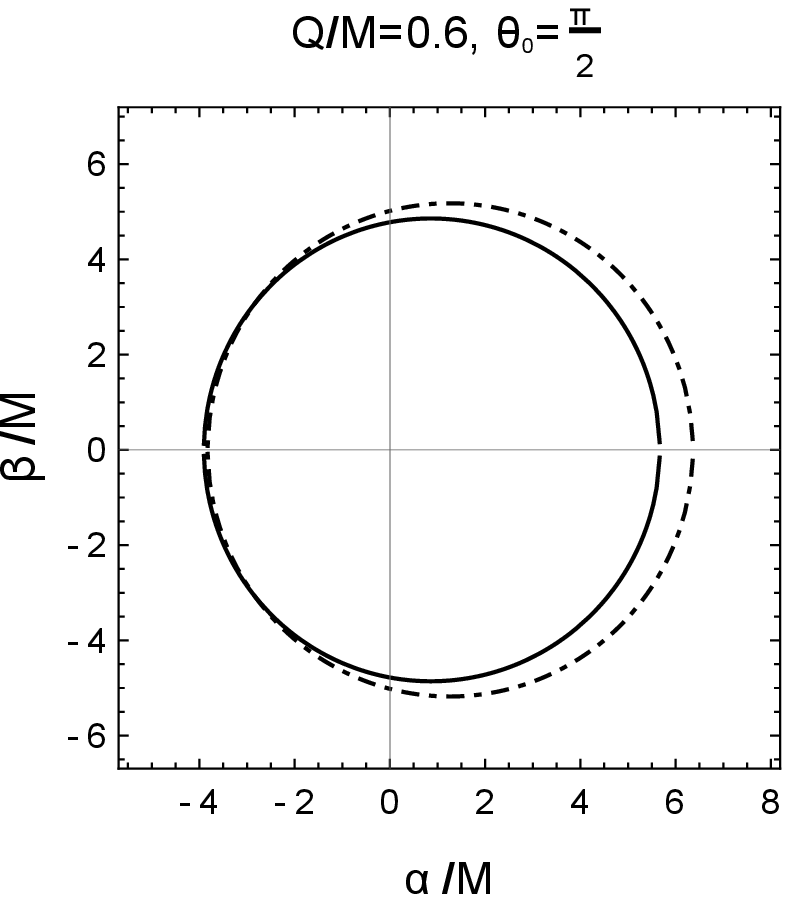}
   \includegraphics[width=.23\textwidth]{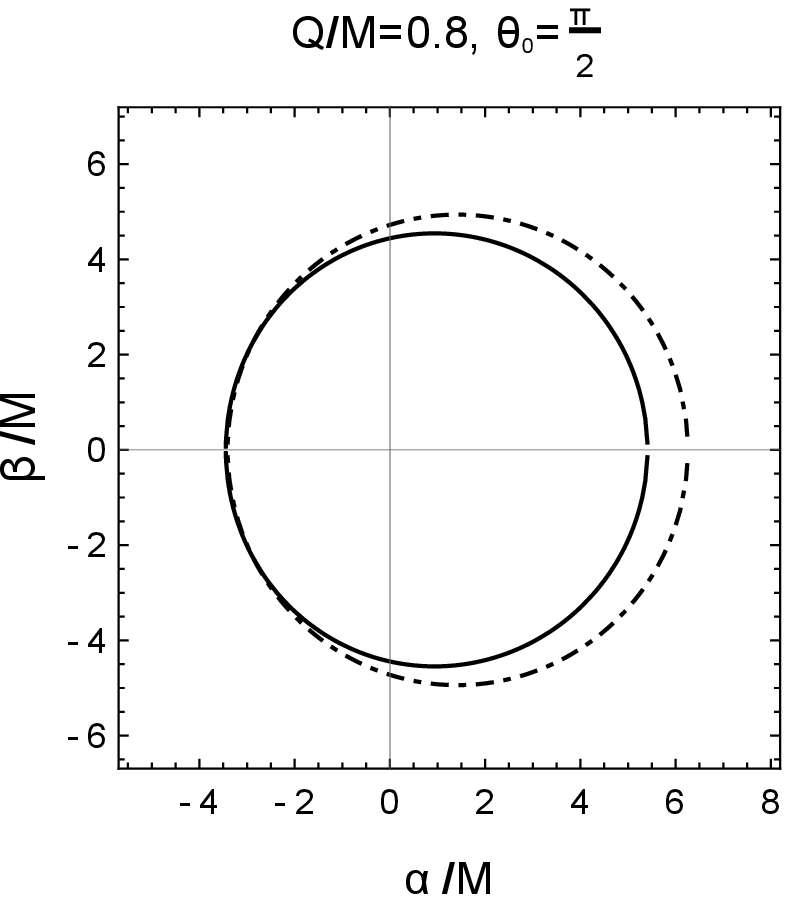}
   \includegraphics[width=.23\textwidth]{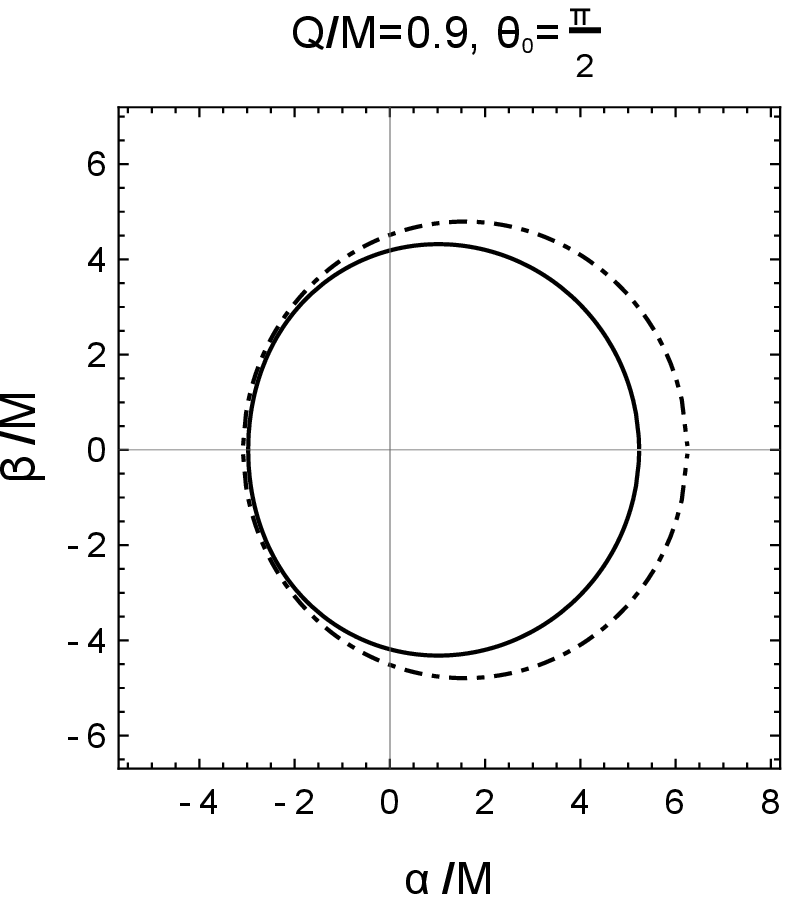}

   \includegraphics[width=.23\textwidth]{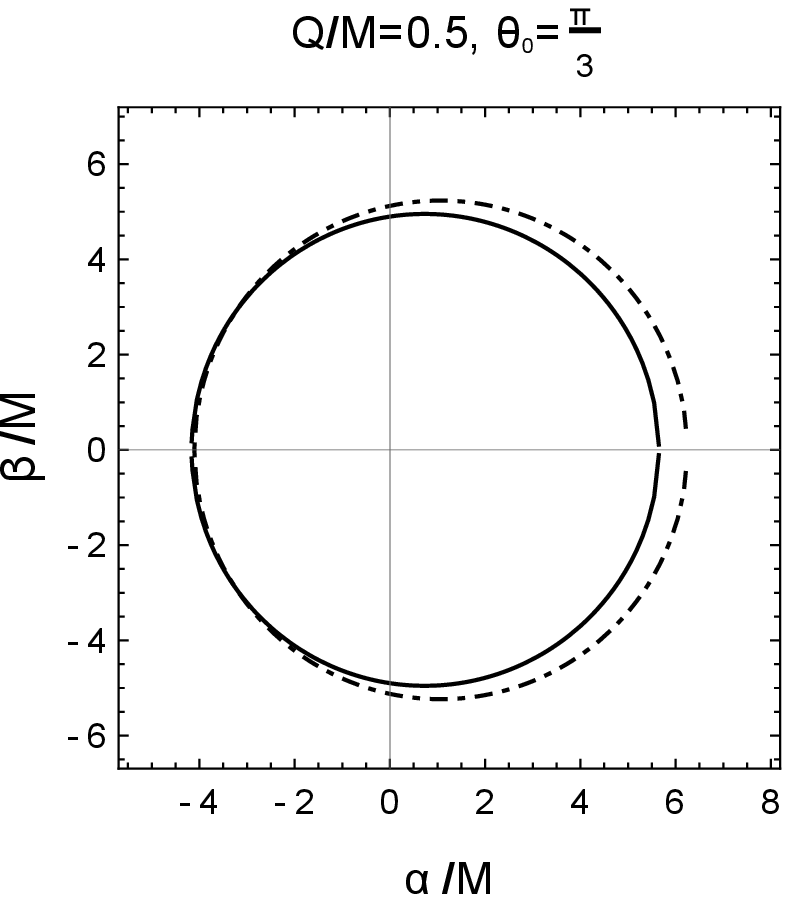}
   \includegraphics[width=.23\textwidth]{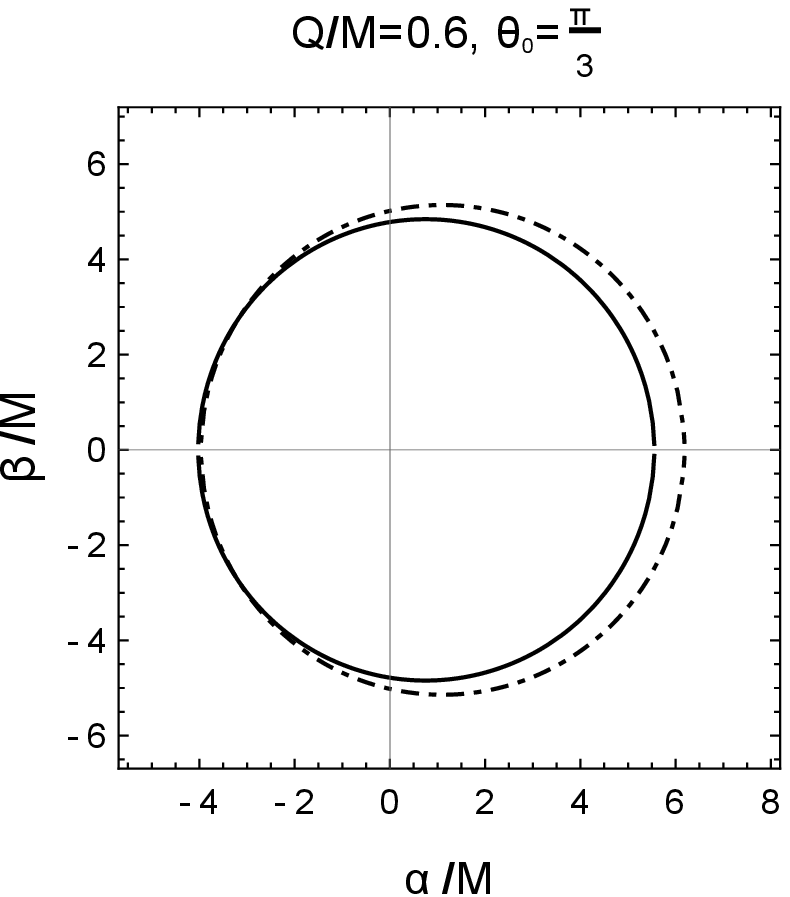}
   \includegraphics[width=.23\textwidth]{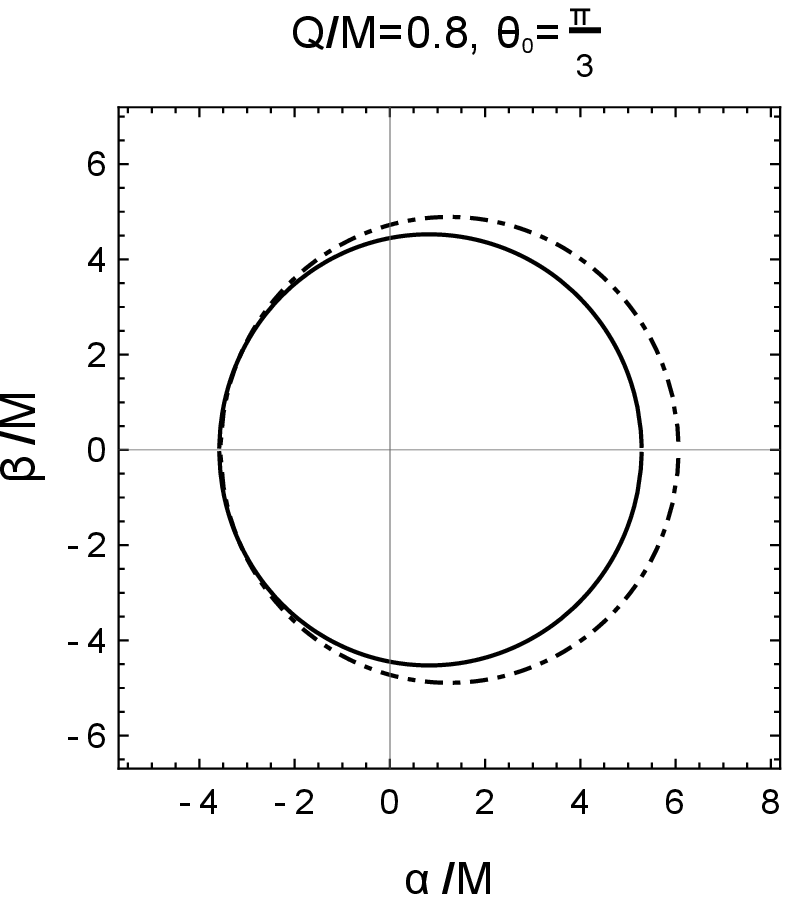}
   \includegraphics[width=.23\textwidth]{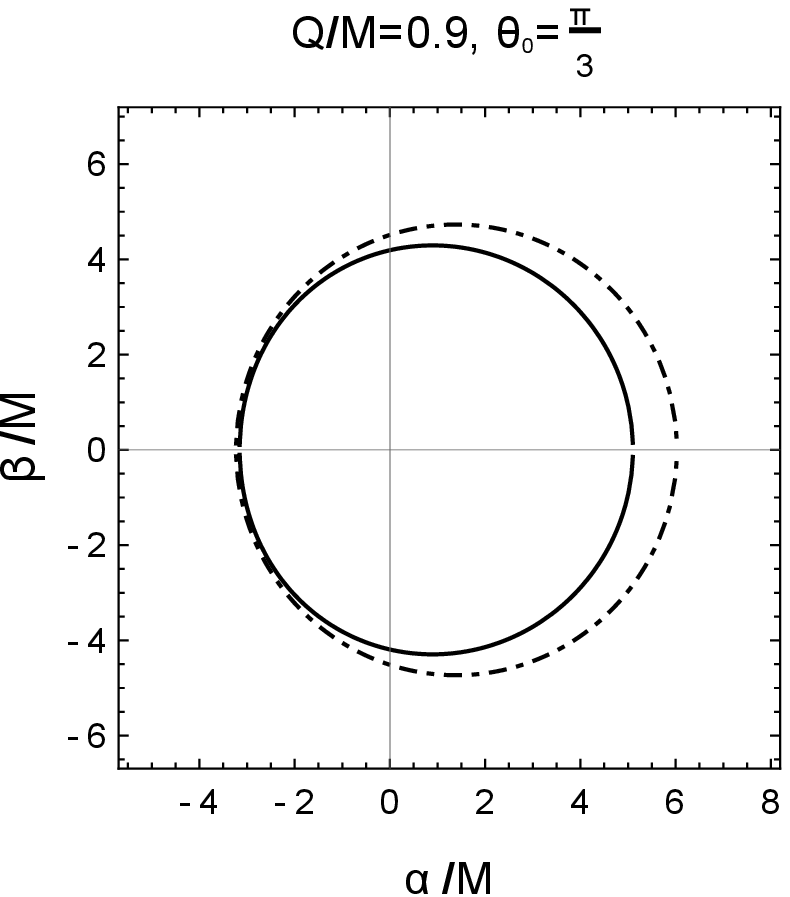}

   \includegraphics[width=.23\textwidth]{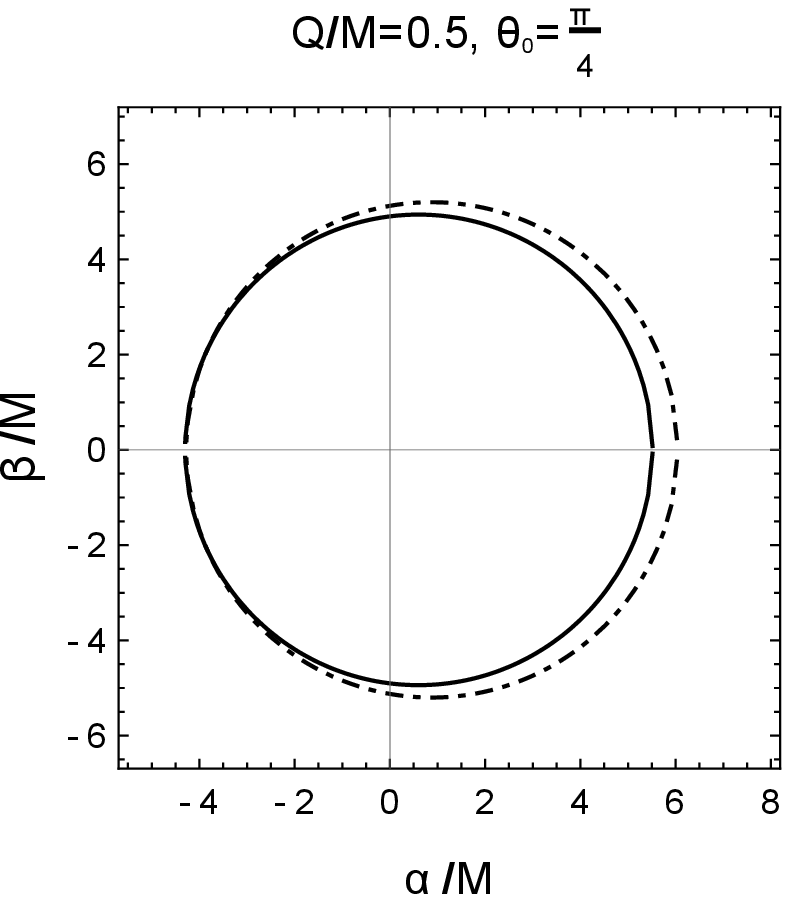}
   \includegraphics[width=.23\textwidth]{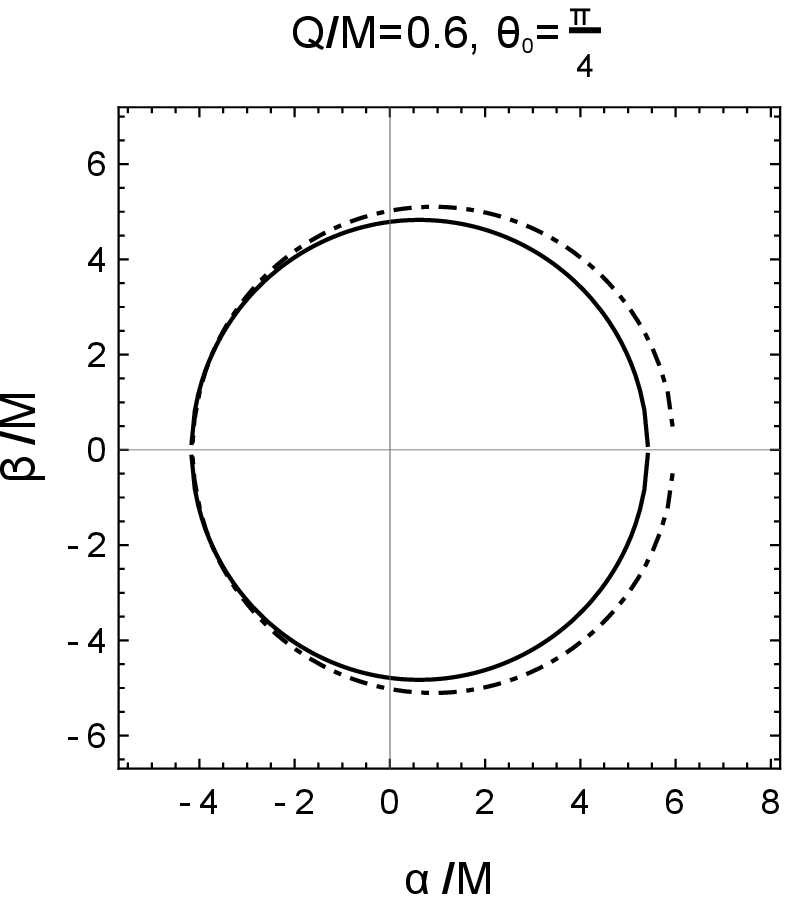}
   \includegraphics[width=.23\textwidth]{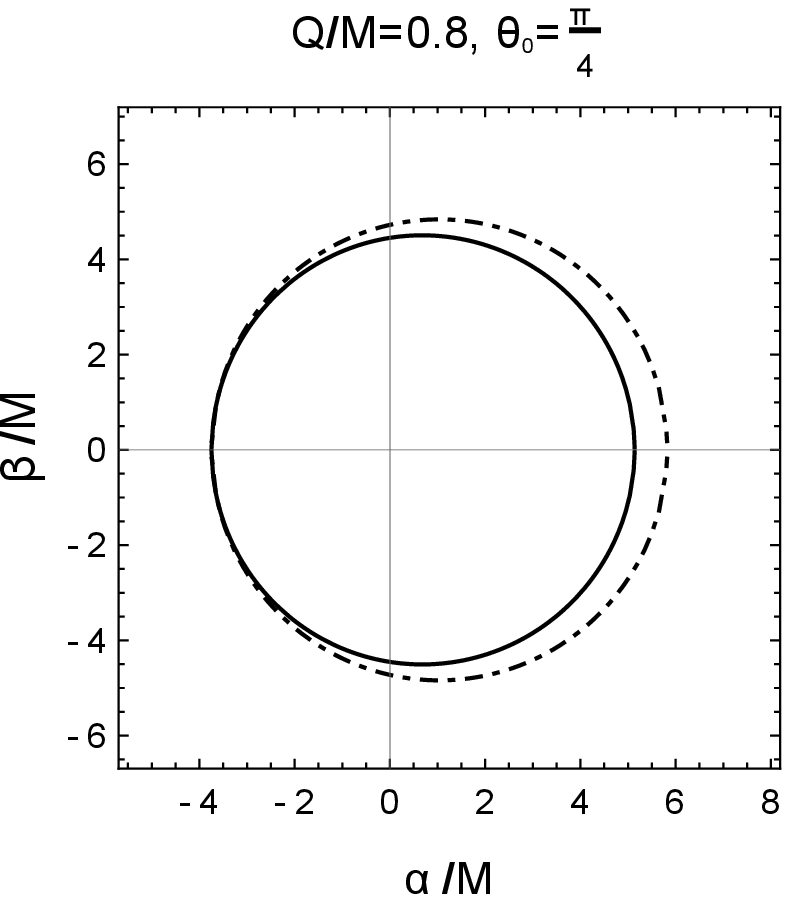}
   \includegraphics[width=.23\textwidth]{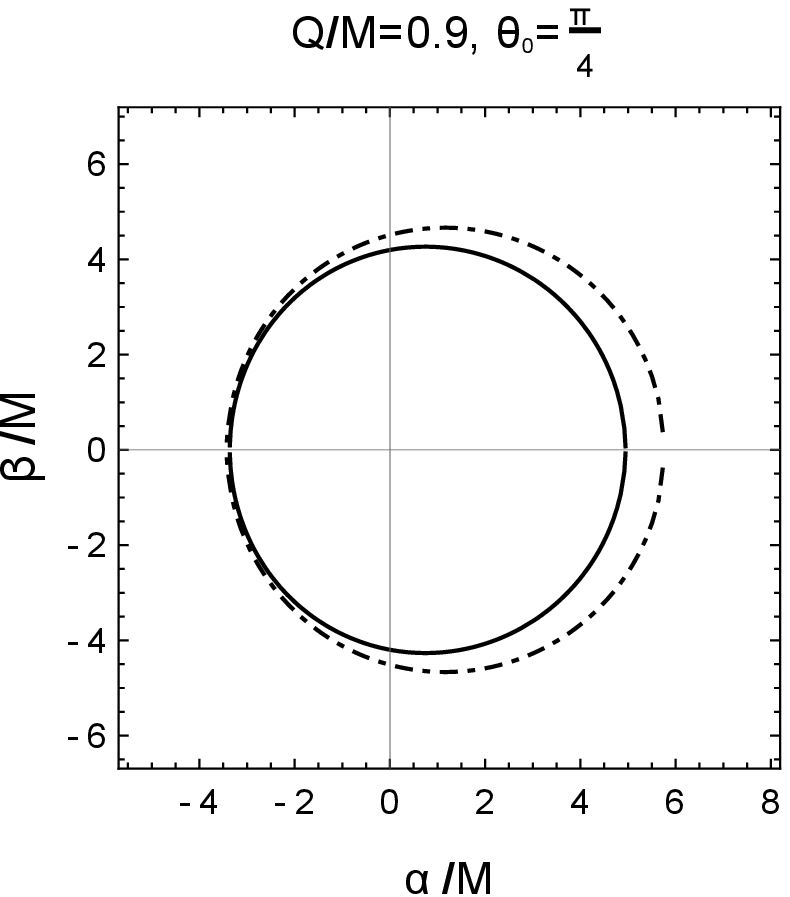}

   \includegraphics[width=.23\textwidth]{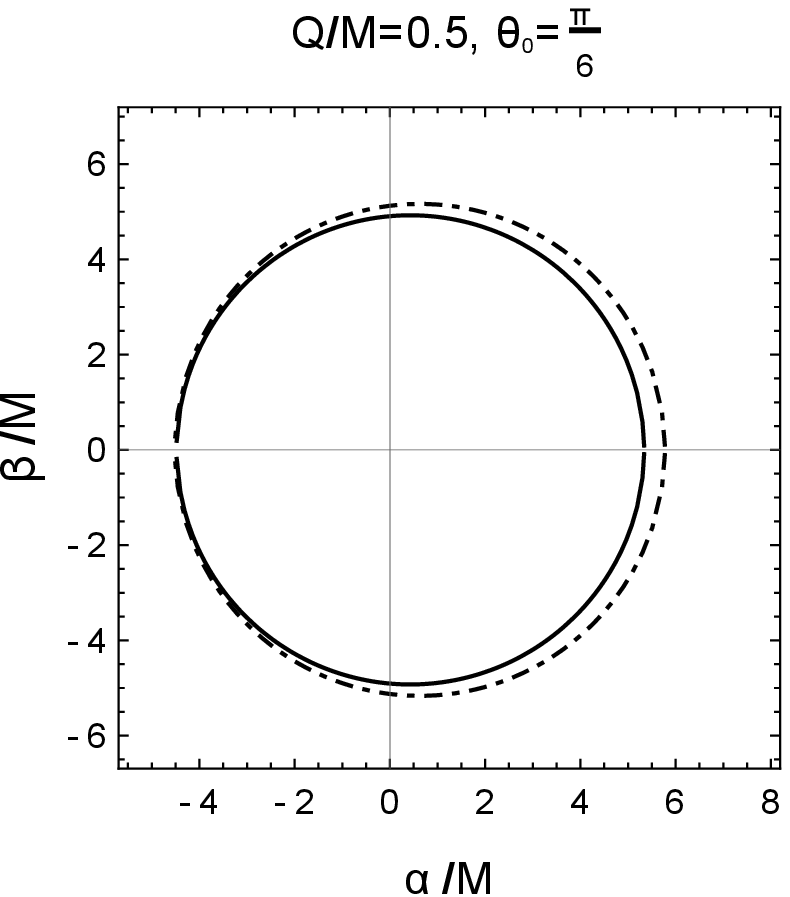}
   \includegraphics[width=.23\textwidth]{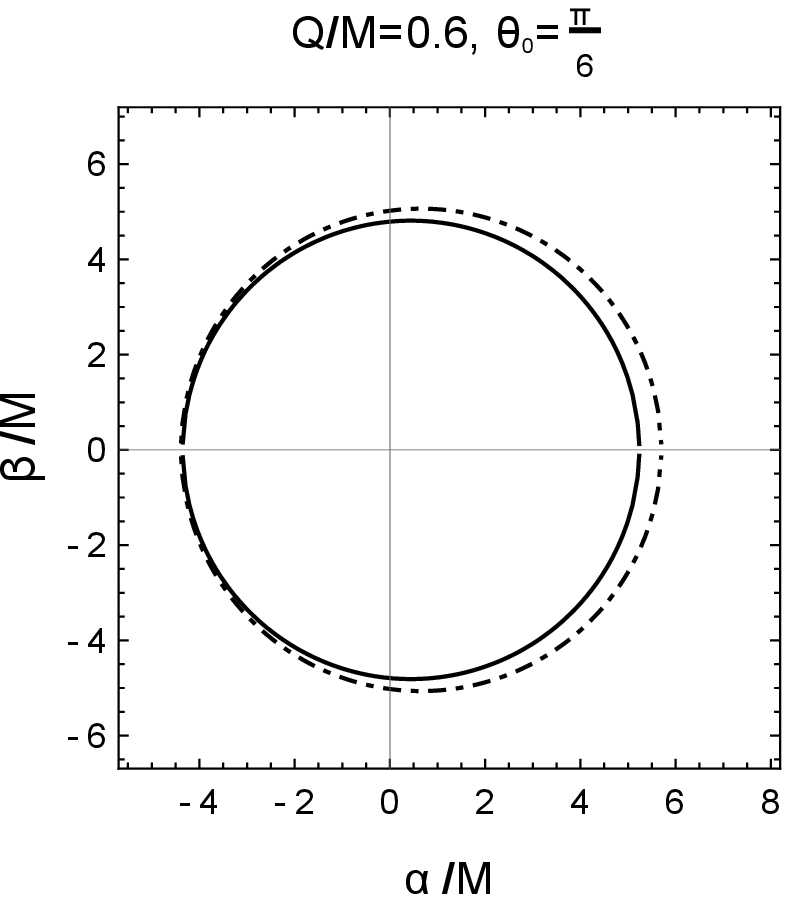}
   \includegraphics[width=.23\textwidth]{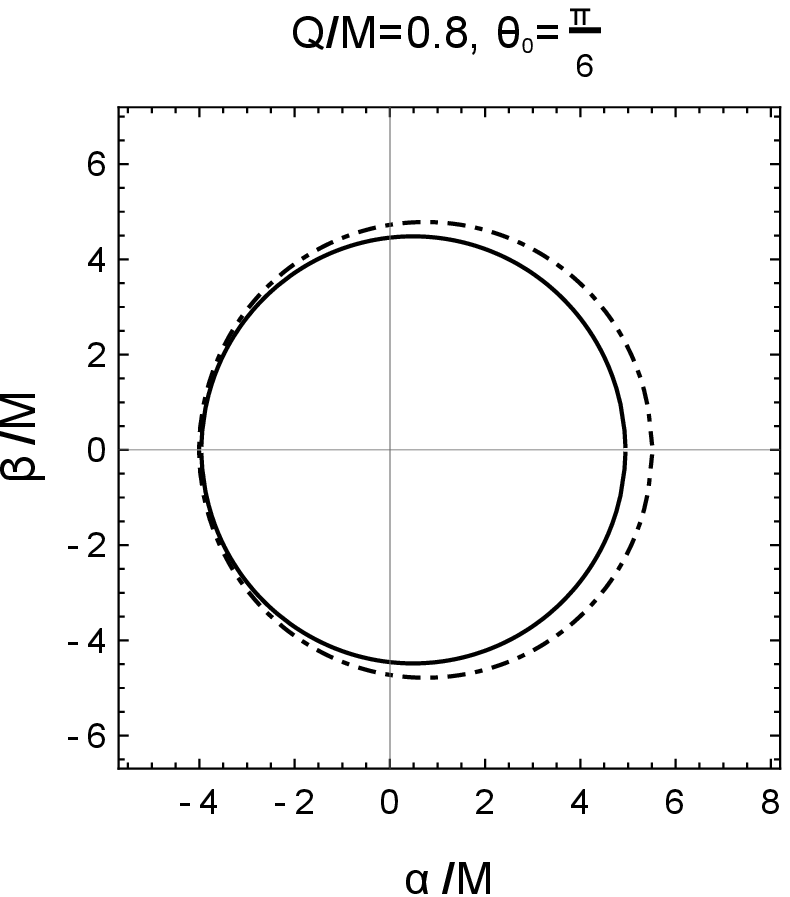}
   \includegraphics[width=.23\textwidth]{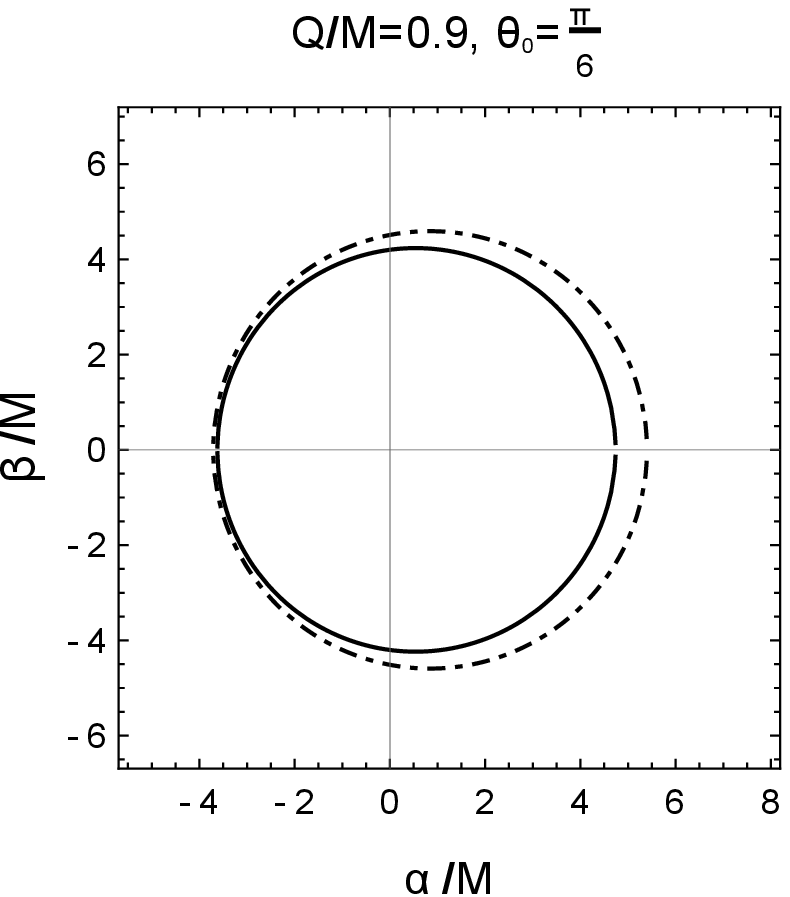}
\end{center}
\caption{The shadow of the black hole surrounded by a plasma for the
 different values of the charge parameter $Q$, inclination angle $\theta_{0}$, and the
 refraction index $n$. The solid lines in the plots correspond to the vacuum case, while for
  dotdashed lines the plasma frequency is $k/M$=0.5 and $a$=0.4.}\label{Shadows}
\end{figure*}

\begin{eqnarray}
\xi&=&A+\sqrt{A^2-B},  \label{xi} \\
\eta&=&\frac{2aM\xi-(a^2+r^2)\big(2rn^2+nn^{'}(a^2+r^2)\big)}{M-r}. \label{eta}
\end{eqnarray}
where,
\begin{eqnarray}
A&=&\frac{\big(M(a^2-r^2)+rQ^2\big)}{a(M-r)},\nonumber \\
B&=&\frac{(a^2+r^2)\big(\big(M(a^2-r^2)+rQ^2+r\Delta\big)n^2+\Delta nn^{'}(a^2+r^2)\big)}{a^2(M-r)}.\nonumber \\
\end{eqnarray}

The contour of the shadow in terms of celestial coordinates, $(\alpha,\beta)$ is given as below,

\begin{eqnarray}
\alpha&=&\lim_{{r_{0}}\rightarrow{\infty}}\bigg(-r_{0}^2 \sin\theta_{0} \frac{d \phi}{d r}\bigg),\\
\beta&=&\lim_{r_{0}\rightarrow{\infty}}\bigg(r_{0}^2\frac{d \theta}{d r}\bigg).
\end{eqnarray}

Calculating $d \phi/d r$ and $d \theta/d r$ using equations (\ref{motionb}-\ref{motiond}), we have

\begin{eqnarray}
\alpha&=& -\frac{\xi \csc\theta}{n},  \label{alpha} \\
\beta&=&\pm\frac{\sqrt{\eta-\xi^2\csc^2\theta-a^2n^2\sin^2\theta}}{n}.\label{beta}
\end{eqnarray}
In Fig. (\ref{Shadows}) the shadow of the rotating black hole for the different values of black hole charge parameter $Q$,
 inclination angle $\theta_{0}$ and the plasma factor is shown. We made a special choice of the plasma frequency in the
 form $\omega_{e}/\omega= k/r$. It is clearly seen in Fig. (\ref{Shadows}), the size and shape of the rotating black hole
 surrounded by the plasma gradually gets modified as the charge and plasma refractive index variate. From a physical perspective
 the change in refractive index occurs as a result of the gravitational red shift phenomenon.

\subsection{Shadow of a non rotating charged black hole in Presence of Plasma}\label{Rh}
Now we focus on the static charged black hole, aiming to understand the charge and plasma effects thoroughly.
Using (\ref{alpha},\ref{beta}) the radius of the static black hole walled in by a plasma is obtained as,
\begin{eqnarray}
 R_{\mathrm{sh}}&=&\frac{1}{n(M-r)}\big[2Mr(Q^2-Mr)-2n^2r^3(M - r) -
 \nonumber \\ && nn^{'} r^4(M -r)+ 2Mr\{\,Q^2(Q^2-2 M r)+M^2r^2-
 \nonumber \\ && n^2r(M - r)\big(2Q^2+r(-3M+r)\big)-
 \nonumber \\ && nn^{'} r^2(M-r)\big(Q^2+r(-2M+r)\big)\,\}^{1/2}\big]^{1/2}\label{radii}
 \end{eqnarray}

\begin{figure*}
 \begin{center}
   \includegraphics[width=.45\textwidth]{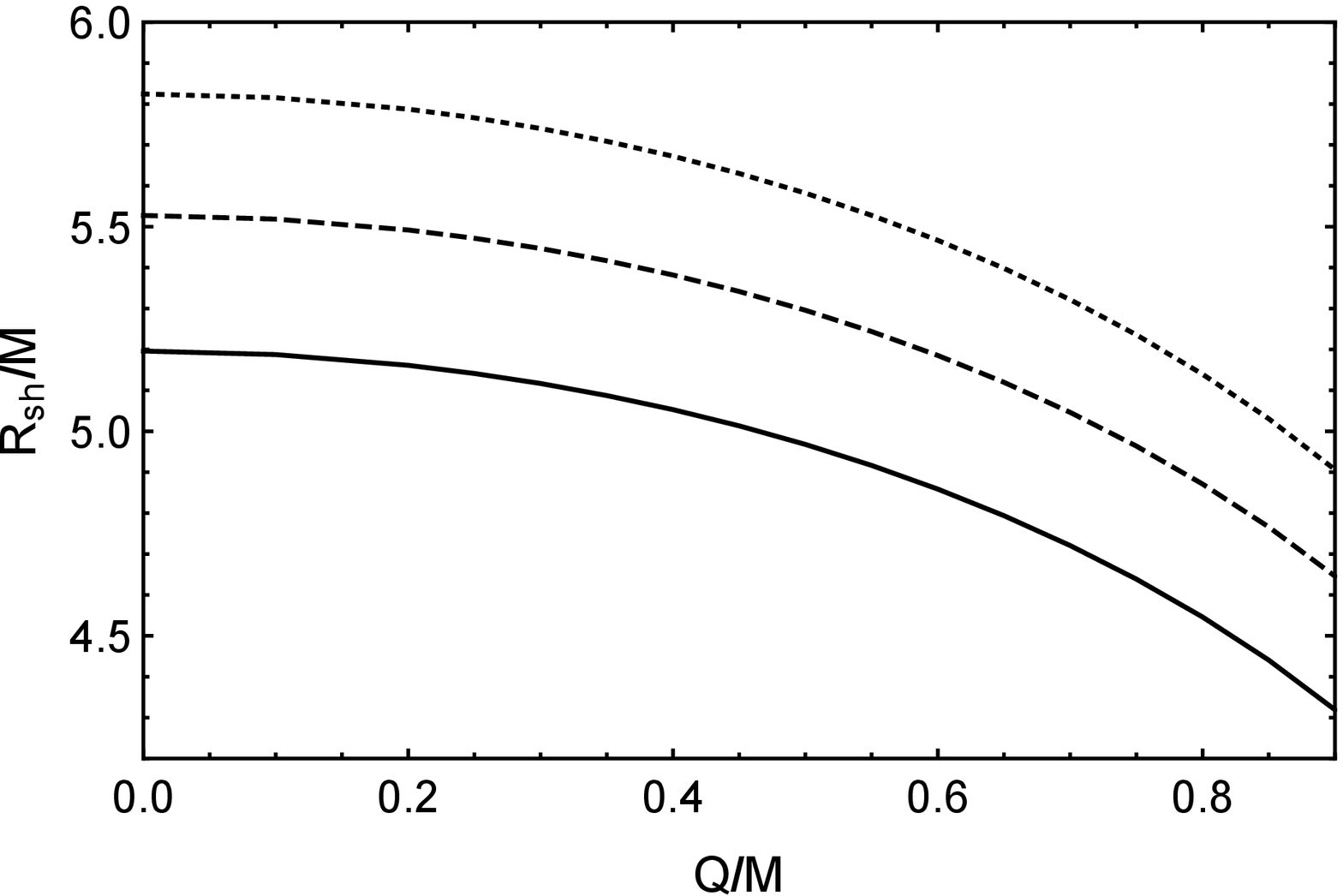}
   \includegraphics[width=.45\textwidth]{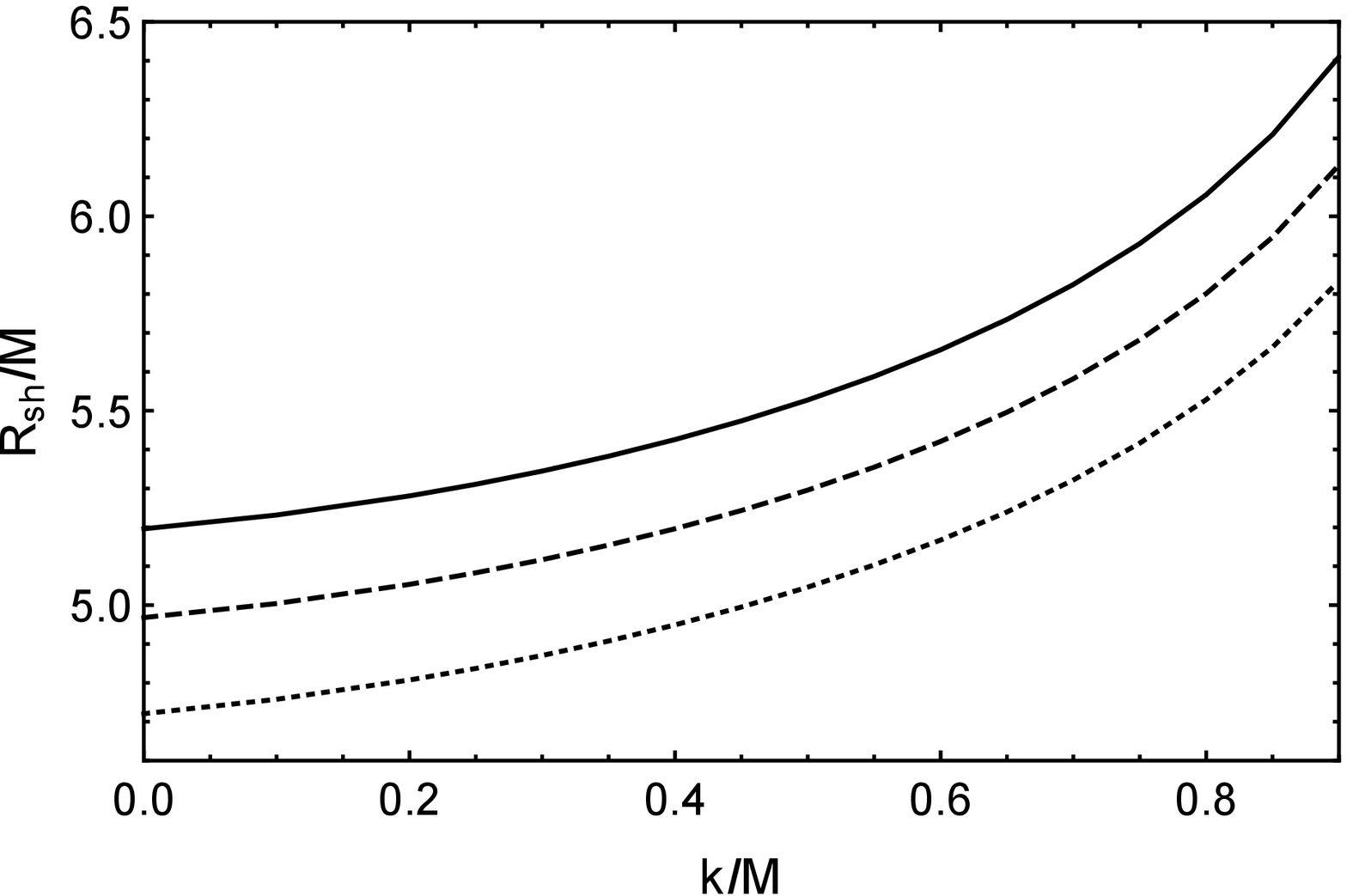}
  \end{center}
\caption{The radius of the shadow as a function of charge and plasma parameter. In the left
panel the solid, dashed and dotted plots correspond to $k/M$= 0,0.5 and 0.7, respectively. In the right panel the solid,
 dashed and dotted plots correspond to $Q/M$=0,0.5, and 0.7, respectively.}\label{Rhadius}
\end{figure*}

\begin{figure*}
 \begin{center}
   \includegraphics[width=.45\textwidth]{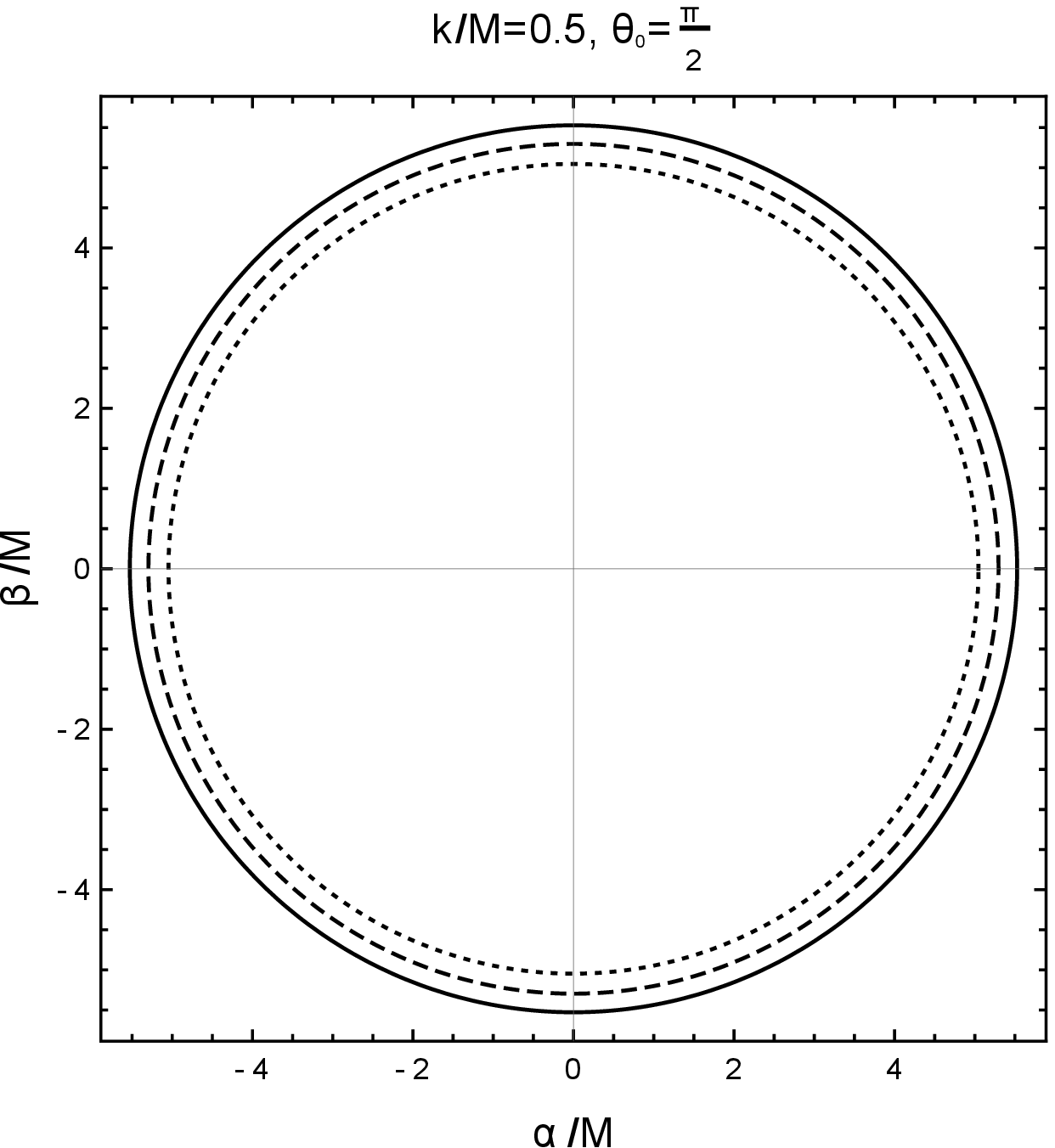}
   \includegraphics[width=.45\textwidth]{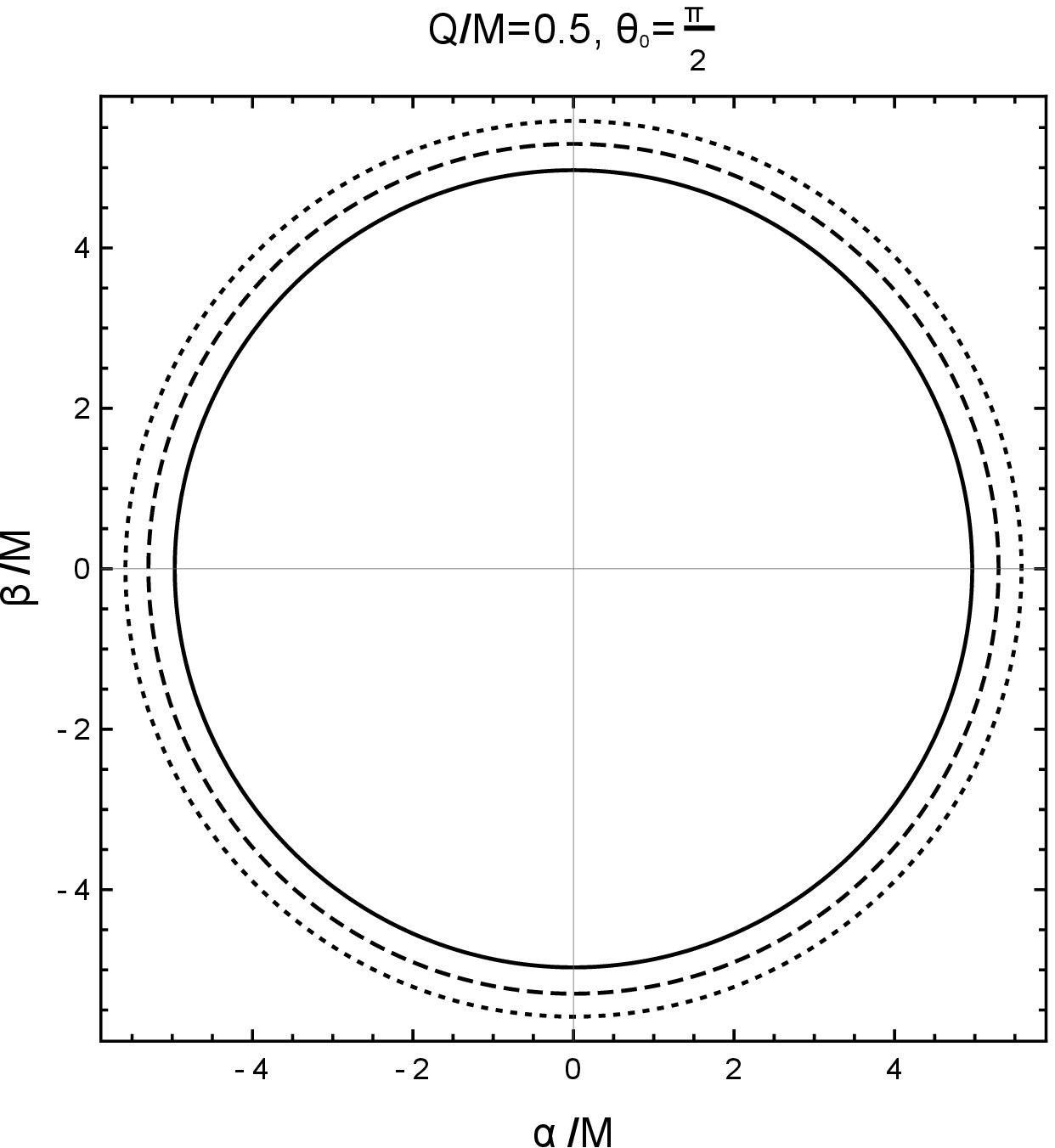}
  \end{center}
\caption{A visualization of the silhouette for distinct charge
 and plasma parameters. In the left panel the solid, dashed and dotted plots correspond
  to $Q/M$=0,0.5 and 0.7, respectively. In the right panel the solid, dashed and dotted plots correspond to $k/M$=0,0.5 and 0.7, respectively.}\label{QP}
\end{figure*}

For $Q=0$, at $r_{\mathrm{ph}}=3M$ we retain the radius of the shadow for Schwarzschild black hole, i.e, $R_{\mathrm{sh}}=3\sqrt{3}M$ \cite{Vir:2000a,Vir:2001b}.
The influence of charge and plasma on the radius of the shadow for a static charged black hole is demonstrated in Fig. (\ref{Rhadius}).
The relative values of $r_\mathrm{ph}$ are obtained numerically using (\ref{photono}).
The results attained are analogous to \cite{Vries:2000a}, the increase in charge makes the shadow
appear smaller to the distant observer while for the plasma parameter we notice contrary effects on the radius. When $k/M=0$, the shadow emerges
in a much smaller size.  In Fig. (\ref{QP}) the evolution of shadow is represented visually with respect to various charge and plasma parameters.
In the left panel, as the value of charge adds up the shadow radius is seen to shrink down.
It is clearly shown in the right panel that the radius becomes larger with the increasing plasma factor.

\subsection{Emission energy of a non rotating charged black hole}\label{En}
Black holes are known to emit thermal radiations which lead to a slow decrease in mass of the black hole until it
completely annihilates \cite{haw:1975a}. We are interested to examine the energy emission of the Kerr-Newman black hole
in the presence of plasma using the relation for the Hawking radiation at the frequency $\omega$

\begin{eqnarray}
\frac{d^2 \mathcal{E}(\omega)}{d \omega d t}&=&\frac{2\pi^2\sigma_{lim}\omega^3}{e^{\frac{\omega}{T}-1}},\label{emission}
\end{eqnarray}

where, $T=\kappa/2\pi$ is the Hawking temperature and $\kappa$ is the surface gravity. Here, for convenience, we consider the
case when the black hole is static and spherically symmetric. The limiting constant $\sigma_{lim}$ defines the value of the absorption cross
section vibration for a spherically symmetric black hole

\begin{eqnarray}
\sigma_{lim}\approx \pi {R_\mathrm{sh}}^2,
\end{eqnarray}

 where, $R_\mathrm{sh}$ is computed from (\ref{radii}). Therefore, (\ref{emission}) takes the form \cite{Wei:2013a}
\begin{eqnarray}
\frac{d^2 \mathcal{E}(\omega)}{d \omega d t}&=&\frac{2\pi^3 {R_\mathrm{sh}}^2 \omega^3}{e^{\frac{\omega}{T}-1}}.
\end{eqnarray}

\begin{figure*}
 \begin{center}
   \includegraphics[width=.4\textwidth]{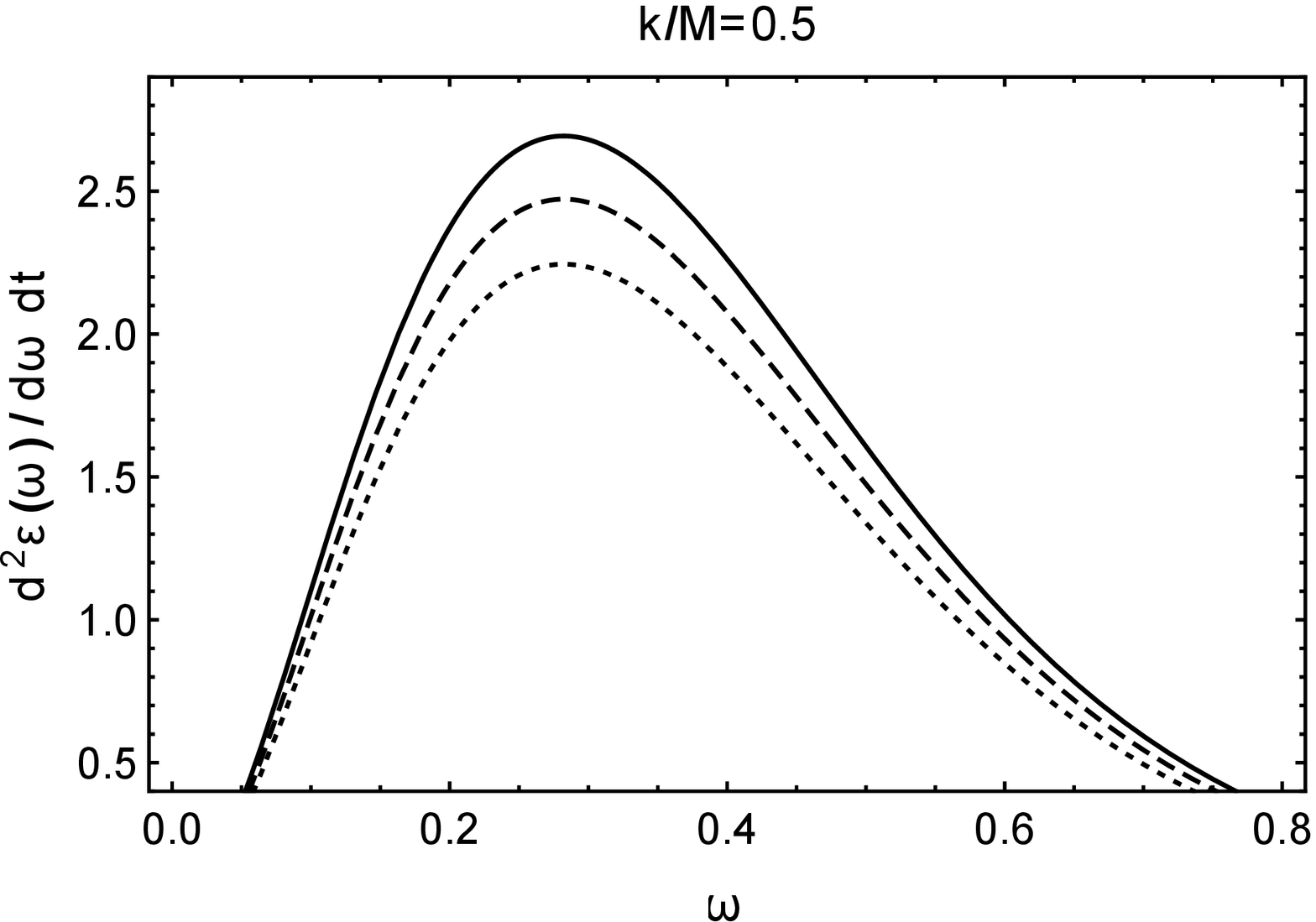}
   \includegraphics[width=.4\textwidth]{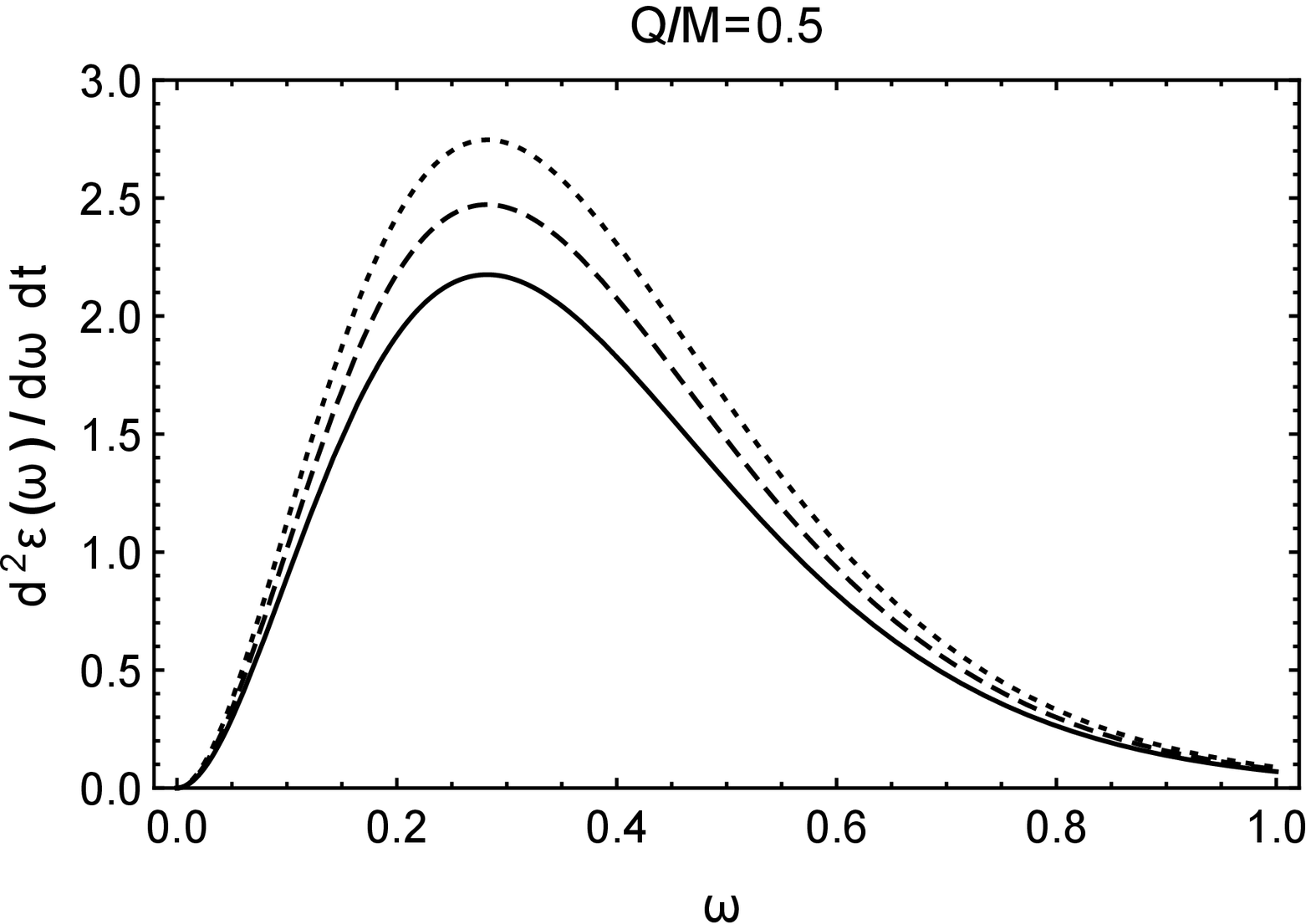}
  \end{center}
\caption{Energy emission of the black hole for distinct charge and plasma parameters. In the left panel
the solid, dashed and dotted plots correspond to $Q/M$=0,0.5 and 0.7, respectively. In the right panel the solid,
 dashed and dotted plots correspond to $k/M$=0,0.5 and 0.7, respectively.}\label{Energy}
\end{figure*}

The energy radiation of a black hole in the presence of plasma is directly proportional to the size of its shadow.
The dependence of the energy emission rate on the frequency for the different values of charge and plasma parameters is
illustrated in Fig. (\ref{Energy}). It is observed that the rate of emission is higher for the small charge values,
thus, at $Q=0$, comparatively a large amount of energy is liberated. In case of the
plasma parameter the emission rate exhibits rise with the increasing
values of the plasma. In the absence of plasma $k/M=0$, a less amount of energy release is detected.

\section{Lensing in Weak Field In the Presence of Plasma}\label{lensing}
We first present a model of weak-field for the non rotating Kerr Newman black hole to
study the gravitational lensing more precisely. The weak-field approximation is given by the relation
\begin{eqnarray}
g_{\mu\nu} = \eta_{\mu\nu} + h_{\mu\nu}\ ,
\end{eqnarray}
$\eta_{\mu\nu}$ and $h_{\mu\nu}$ refer to the Minkowski metric and perturbation metric, respectively. They satisfy the below mentioned properties
\begin{eqnarray}
&&\eta_{\mu\nu} = {\rm diag} (-1, 1, 1, 1)\ ,\nonumber\\
&& h_{\mu\nu} \ll 1, \quad  h_{\mu\nu} \rightarrow 0 \quad {\rm under } \quad x^i\rightarrow \infty\ , \nonumber \\
&& g^{\mu\nu}=\eta^{\mu\nu}-h^{\mu\nu},\ \ \ h^{\mu\nu}=h_{\mu\nu}\ .
\end{eqnarray}

Taking into account the weak-field approximation and weak plasma strength, for photon propagation
along $z$ direction, one can easily obtain the angle of deflection complying the steps in \cite{Abu:2017aa}, we have

\begin{eqnarray}\label{alphak}
&&\hat{\alpha}_k= \frac{1}{2} \int_{-\infty}^\infty \left(h_{33} +\frac{h_{00} \omega^2-{\omega_{e}}^2}{\omega^2-\omega_e^2}\right)_{,k} dz \ .\nonumber \\
\end{eqnarray}

Note that negative and positive sign for  $\hat{\alpha}_b$ indicate respectively deflection towards and away from the central object. At large $r$, the black hole metric could be approximated to \cite{Abu:2017a}
\begin{eqnarray}
ds^2 = ds^2_{0}+\left(\frac{2M}{r}-\frac{Q^2}{r^2}\right)dt^2+\left(\frac{2M}{r}-\frac{Q^2}{r^2}\right) dr^2  , \label{metr1}
\end{eqnarray}

where
$ds^2_{0}=-dt^2+dr^2+r^2(d\theta^2+\sin^2\theta d\phi^2)$.

In the Cartesian coordinates the components $h_{\mu\nu}$ can be written as
\begin{eqnarray}
 h_{00}&=&\left(\frac{R_g}{r}-\frac{Q^2}{r^2}\right), \nonumber \\  h_{ik}&=&\left(\frac{R_g}{r}-\frac{Q^2}{r^2}\right)n_{i}n_{k}\ ,\nonumber \\
 h_{33}&=&\left(\frac{R_g}{r}-\frac{Q^2}{r^2}\right)\cos^2\chi \ ,
\end{eqnarray}\label{h}

where $R_g=2M$.

Using the above expressions in the formula (\ref{alphak}), one can compute the light deflection angle for a black hole in plasma
\begin{eqnarray}
\hat{\alpha}_{b}&=&\int_0^{\infty}\frac{\partial}{\partial b}\Bigg[ \left(\frac{R_g}{\sqrt{b^2+z^2}}-\frac{Q^2}{b^2+z^2}\right)
\frac{z^2}{b^2+z^2}\
 \nonumber \\ && +\frac{1}{1-\omega^2_{e}/\omega^2}\left(\frac{R_g}{\sqrt{b^2+z^2}}
-\frac{Q^2}{b^2+z^2}\right)\Bigg]dz\ ,
\label{alfa}
\end{eqnarray}

where $b^2=x_1^2+x_2^2$ is the impact parameter, and $x_1$ and $x_2$ are the coordinates on the plane orthogonal to the $z$ axis, and the photon frequency at large $r$ is given by

\begin{eqnarray}\label{omeg1}
 \omega^2=\frac{\omega_{\infty}^2}{(1-\frac{R_g}{r}+\frac{Q^2}{r^2})} .
\end{eqnarray}
Here, $\omega_\infty$ is the asymptotic value of photon frequency. In the approximation of the charge and large distance, the expression (\ref{refractiveindex}), {after expanding in series on the powers of $1/r$,} can be approximated to
%
%
\begin{eqnarray}
n^2=\left({1-\frac{\omega^2_{e}}{\omega^2}}\right)^{-1} \simeq 1+\frac{4\pi e^2 N_0 r_0}{m \omega_\infty^2 r}-\frac{4\pi e^2 N_0 r_0 R_g}{m \omega_\infty^2 r^2}\ .
\end{eqnarray}\label{omega}
Using this approximation one can easily find the deflection angle $\hat{\alpha}_{b}$ of the light around a black hole in presence of plasma
\begin{eqnarray}\label{alfabrane}
\hat{\alpha}_{b} &=&\frac{2R_g}{b}\left(1+\frac{\pi^2 e^2 N_0 r_0}{m \omega_\infty^2 b}-\frac{4\pi e^2 N_0 r_0 R_g}{m \omega_\infty^2 b^2}\right)\nonumber \\ && -\frac{Q^2}{4b^2}\left(3\pi+\frac{4\pi e^2 N_0 r_0}{m \omega_\infty^2 b}\left(8-\frac{3\pi R_g}{b}\right)\right)\ .
\end{eqnarray}

\begin{figure*}
 \begin{center}
   \includegraphics[scale=0.42]{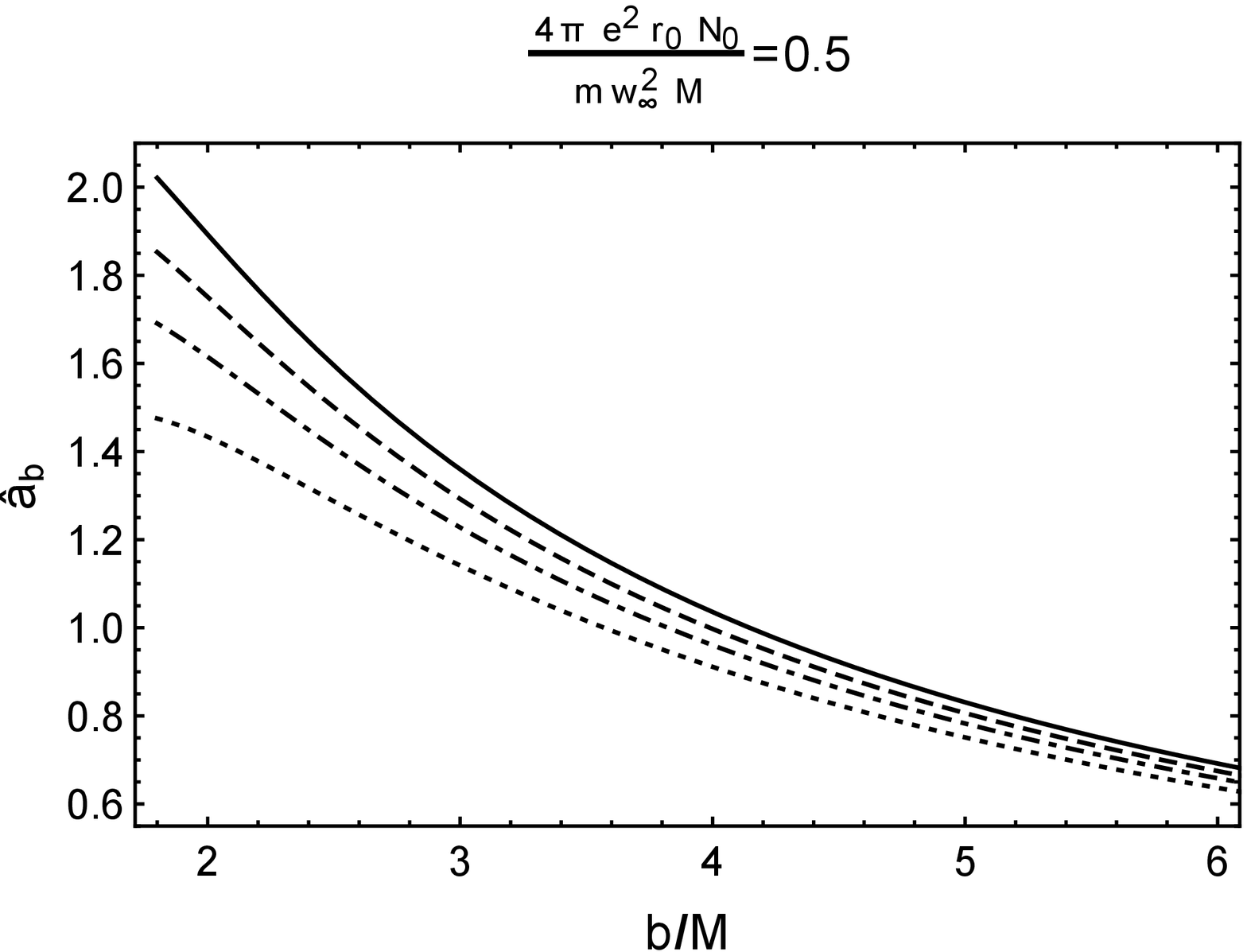}
   \includegraphics[scale=0.42]{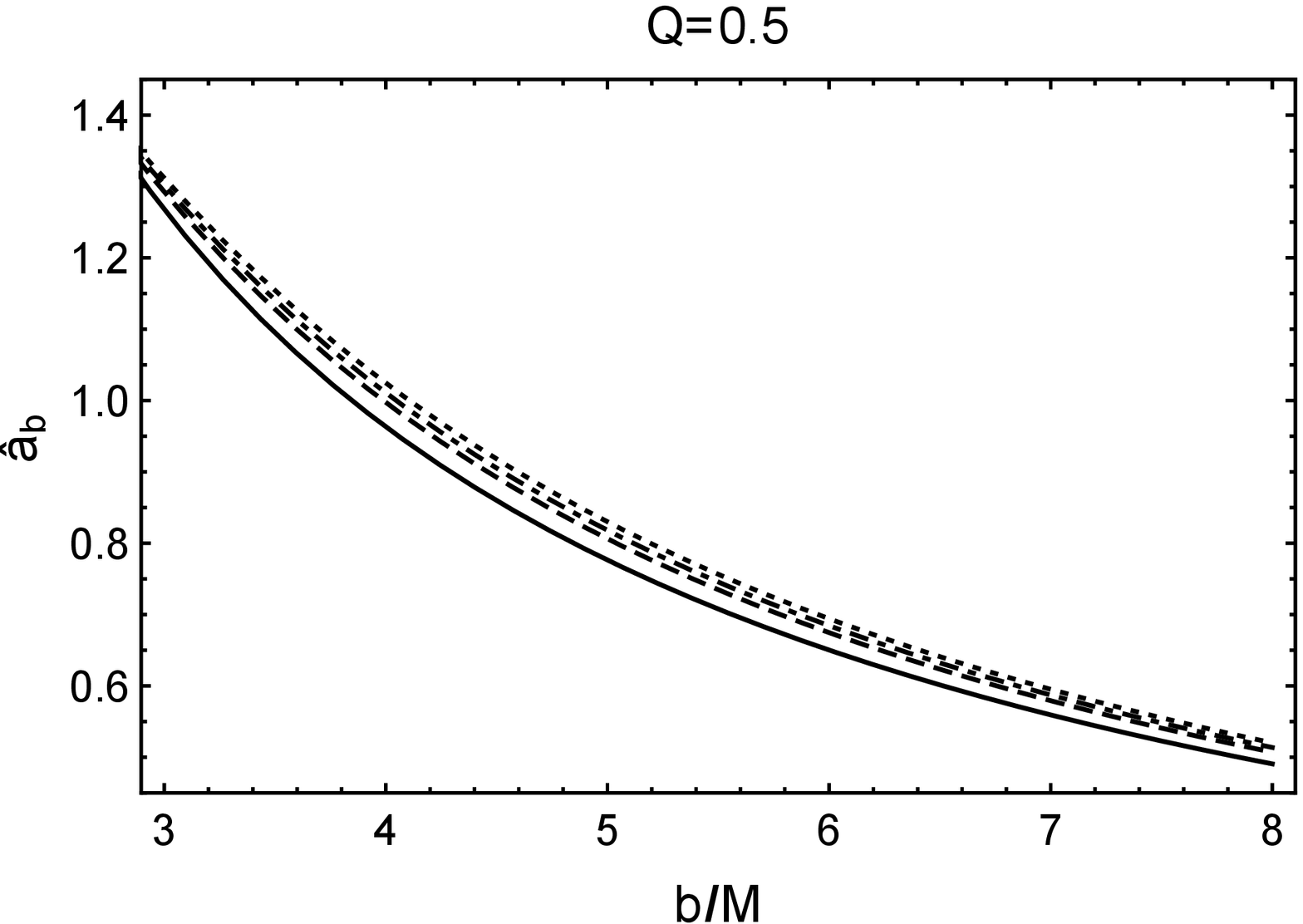}
  \end{center}
\caption{Deflection angle $\hat{\alpha}_{b}$ as a function of the impact
parameter $b$ for different charge and plasma values. In the left panel the values of $Q$ (top to bottom) are 0,0.5,0.7 and 0.9 and
in the right panel the values $\frac{4\pi e^2 N_0 r_0}{m \omega_\infty^2 M}$ (bottom to top) are 0,0.5,0.7 and 0.9.}\label{angle}
\end{figure*}

We get $\hat{\alpha_b} =2R_g/b$ in the absence of charge and plasma for the Schwarzschild black
hole \cite{Tsp:2015a}. The dependence of the angle of deflection $\hat{\alpha_b}$ on the impact parameter $b$ for various charge and plasma
parameters is demonstrated in Fig. (\ref{angle}). In the left panel as the value of charge increases the angle of deflection decreases
and it is seen that $\hat{\alpha_b}$ is maximum when the charge
is turned off, i.e., $Q=0$. We observe that the deflection angle $\hat{\alpha_b}$ increases
with the gradual supplement in the plasma parameter (right panel). Also, one can
clearly notice that the deviation of photons is smaller when the plasma factor is removed from the black hole background.

\section{Conclusion} \label{con}
In this paper we reviewed some well known features of the black hole, i.e.,
the black hole silhouette, energy emission and the weak field lensing in the background of the Kerr-Newman gravity walled in by a plasma medium.
The results are recovered for the Schwarzschild metric when the charge and spin parameter are excluded. The impact of the charge and plasma parameters
on the aforesaid properties have been investigated explicitly. It is construed that the shadow size viewed by
a distant observer is smaller as the charge parameter is increased and by supplementing the plasma factor the size of the shadow appears to be larger.
Since, the energy liberated from the black hole depends on the radius of the shadow, therefore, rate of energy emission from the black hole is higher
when the black hole is surrounded by a plasma. As far as angle of deflection is concerned, the photons are observed
 to experience an increase in the deviation as the plasma factor gradually adds up.
While on the other hand, angle of deflection sufficiently reduces when the amount of charge parameter rises. Nevertheless, it is
investigated that the Kerr-Newman black hole experiences a contradictory influence of the charge and plasma parameters.

\section*{Acknowledgements} F.A. was supported by Grants No. VA-FA-F-
2-008, No.MRB-AN-2019-29 of the Uzbekistan Ministry for Innovative Development and INHA University in Tashkent.

\bibliographystyle{Plasma}
\bibliography{Plasma}

\end{document}